\documentclass{emulateapj}

\newcommand{\gcc}{\ \mathrm{g\ cm^{-3} }}

\newcommand{\Lewis}{ \mathrm{Le} }

\newcommand{\Karlovitz}{ \mathrm{Ka} }
\newcommand{\Markstein}{ \mathrm{Ma} }
\newcommand{\MarksteinLength}{ {\cal{L}_M} }
\newcommand{\Zeldovich}{ \mathrm{Ze} }

\slugcomment{Accepted to ApJ}

\newcommand{\FLASH}{{\sc{Flash}}~}
\newcommand{\enuc}{\dot \epsilon_{\mathrm{nuc}}}
\newcommand{\avg}[1]{\left<#1\right>}
\newcommand{\avgsub}[2]{\left<#1\right>_{#2}}

\usepackage{epsf,color,dcolumn}

\newcolumntype{d}{D{.}{.}{-1}}

\epsfverbosetrue

\begin{document} 

\title{The Response of Model and Astrophysical Thermonuclear Flames to Curvature and Stretch}

\shorttitle{Flame Response to Curvature and Strain}
\shortauthors{Dursi, Zingale, et al.}

\author{
	L.~J.~Dursi\altaffilmark{1,2},
        M.~Zingale\altaffilmark{3}, 
        A.~C.~Calder\altaffilmark{1,2},
        B.~Fryxell\altaffilmark{2},
        F.~X.~Timmes\altaffilmark{1,2}, 
        N.~Vladimirova\altaffilmark{1,2},
        R.~Rosner\altaffilmark{1,2,4},
        A.~Caceres\altaffilmark{2,4},
        D.~Q.~Lamb\altaffilmark{1,2},
        K.~Olson\altaffilmark{2,5},
        P.~M.~Ricker\altaffilmark{6},
	K.~Riley\altaffilmark{2},
        A.~Siegel\altaffilmark{2},
        J.~W.~Truran\altaffilmark{1,2}}

\altaffiltext{1}{Dept.\ of Astronomy \& Astrophysics, 
                 The University of Chicago, 
                 Chicago, IL  60637}

\altaffiltext{2}{Center for Astrophysical Thermonuclear Flashes, 
                 The University of Chicago, 
                 Chicago, IL  60637}
             
\altaffiltext{3}{Dept.\ of Astronomy \& Astrophysics,
                 The University of California, Santa Cruz,
                 Santa Cruz, CA 95064}

\altaffiltext{4}{Dept.\ of Physics, 
                 The University of Chicago, 
                 Chicago, IL  60637}

\altaffiltext{5}{UMBC/GEST Center, NASA/GSFC, 
                 Greenbelt, MD 20771}

\altaffiltext{6}{Dept. of Physics, University of Illinois, 
                 Urbana-Champaign, IL 61801}

\begin{abstract}

Critically understanding the `standard candle'-like behavior of Type
Ia supernovae requires understanding their explosion mechanism.  One
family of models for Type Ia Supernovae begins with a deflagration in
a Carbon-Oxygen white dwarf which greatly accelerates through wrinkling
and flame instabilities.  While the planar speed and behavior of
astrophysically-relevant flames is increasingly well understood, more
complex behavior, such as the flame's response to stretch and
curvature, has not been extensively explored in the astrophysical
literature; this behavior can greatly enhance or suppress instabilities
and local flame-wrinkling, which in turn can increase or decrease the
bulk burning rate.  In this paper, we explore the effects of curvature
on both nuclear flames and simpler model flames to understand the
effect of curvature on the flame structure and speed.
\end{abstract}


\section{INTRODUCTION}
\label{sec:intro}
Type Ia supernovae are used as `standard candles' for cosmology (see the
review by \citealt{hillebrandtniemeyer00} and references therein).  The
standard model for a Type Ia supernova involves a flame that begins
deep in the interior of a Carbon-Oxygen Chandrasekhar mass white dwarf,
but the full mechanism for explosion is not yet understood.
One-dimensional simulations (\citealt{woosley84,nomoto84}) of
flame-powered Type Ia can match the energetics of the observations if
the flame reaches $\sim$1/3 of the speed of sound as it burns through
the star, although the precise mechanism for this flame acceleration is
still poorly understood.  Other models begin with a flame that
accelerates to $\sim$1/30 of the sound speed \citep{dominguez00} and
undergoes an as-yet unexplained transition to a detonation
\citep{khokhlov91a,khokhlov91b,arnett94a,arnett94b}.  Either mode
requires that the flame accelerate considerably beyond its laminar
speed, and flame instabilities are generally cited as potential
mechanisms.  When burning begins, pockets of buoyant, hot ash are
formed as the flame propagates outward.  This flame is unstable to the
Landau-Darrieus and Rayleigh-Taylor instabilities \citep{COld}.  These
instabilities wrinkle the flame front, increasing the surface area, and
therefore, the bulk burning rate.

Both strain from curvature and flow-induced stretch of a flame are
known to affect a flame's speed and structure \citep{markstein64}.
These strains change both the local burning rate, and the bulk burning
behavior in complex large scale flows \citep{ldlargescale} through
multidimensional flame instabilities.  Curvature effects in terrestrial
flames have received a great amount of attention (see, for instance,
the review by \citealt{lawsung}), but the astrophysical flame literature
on these effects has been sparse.

The instabilities that might increase the total burning by a flame do
so by stretching and curving the flame.  Because these strains
themselves can significantly change a flame's local burning rate, the
onset and growth rate of flame instabilities are directly affected by
curvature.  Thus, in the context of the standard Type Ia model,
understanding the detailed explosion mechanism requires
understanding the micro-physics of the flame propagation in the white dwarf.
One way of doing this would be to extend large supernova simulations
(eg,
\citealt{mpa3dsn1a,mpa2dsn1a,niemeyer2dsn1,gamezo03}) to incorporate a fully resolved flame. However, at the
onset of the burning, the flame is only $\sim 10^{-5} \ \mathrm{cm}$
thick, inside a star of radius $\sim 10^8 \ \mathrm{cm}$.  At lower
densities, the flame becomes broader, but remains tiny compared to the
radius of the whole star.  Thus any realistic study of
multidimensional Type Ia scenarios that follow a significant portion of
the star requires understanding of the behavior of the flames, and
must use a model on unresolved scales to describe the burning physics,
such as flame-stretch interactions and the corresponding effects on
instabilities.  Some such models exist \citep{khoklov95,MPA02}, but
they are based on the properties of large scale flows and terrestrial
scalings, rather than on the local physical properties of astrophysical
flames.  A detailed understanding of the flame's response to curvature
is necessary input to a subgrid scale flame model.

In this paper, we describe numerical experiments of propagating,
spherical flames, similar to recent experiments and calculations
performed by the chemical combustion community \citep{bradley96,
karpov97, aung97, hassan98, bradley98, sun99, gu00}, designed to measure
the flame structure and speed response to geometric curvature.  All of
the flame simulations presented are 1-d (either in planar or spherical
geometry), to get clean and unambiguous measurements of the flame's
response to curvature.   As the flame propagates to different radii, it
experiences different strain due to the geometry, parameterized by a
dimensionless strain rate, $\Karlovitz$, defined in later sections.  
Quantitatively, we seek the modified flame velocity 
\begin{equation} 
S_l = S_l^0 \cdot f(\Karlovitz),
\end{equation}
where $S_l$ is the laminar flame speed, and $S_l^0$ is the uncurved
laminar flame speed.  (Throughout this paper, a superscript 0 will refer
to planar, unstretched flame quantities.)

\cite{markstein64} postulated a flame-speed law of the above form for
small curvatures based on empirical results.  Expressed in dimensionless
numbers, this is written:
\begin{equation} 
S_l = S_l^0 \cdot \left ( 1 + \Markstein \, \Karlovitz \right ) ,
\label{eq:marksteindef} 
\end{equation}
where $\Markstein$ is the dimensionless Markstein number, which depends
on parameters of the flame.  We refer to this as the Markstein relation.
The Markstein number can be negative or positive for any given flame.
A negative Markstein number indicates that the flame would burn more
slowly in regions of positive curvature ($\Karlovitz > 0$) and more
quickly in regions of negative curvature ($\Karlovitz < 0$), and the
reverse is true for a flame with positive Markstein number.

The Markstein relation has proven very robust for describing flame
behavior in terrestrial flames.   In that context, work has been done to
calculate Markstein numbers from first principles (for example, \citealt{
clavinwilliams, matalonmatkowsky, chunglaw88}) and from numerical or
physical experiments (for example, \citealt{karpov97,muller97,kwon01,
aung98,gu00,hassan98,bradley96,bradley98,sun99}).  In this paper,
we discuss its applicability to astrophysical thermonuclear flames
and examine the flame-curvature relation both for these flames and for
simpler model flames with parameters that may be varied to understand
their effects.

We outline some theory relevant to our study in \S\ref{sec:theory}.  In
\S\ref{sec:methods} we discuss the methods used in the numerical
experiments.  These experiments, which are described in
\S\ref{sec:experiments}, were performed to understand how the flame
speed changes with curvature.  In \S\ref{sec:results} we present the
results of these experiments, and we conclude in
\S\ref{sec:conclusions}.  The appendix provides some simple tests that
provide verification of our code for the present simulations.

\section{THEORY}
\label{sec:theory}
\subsection{Planar Astrophysical Flame Structure}

\begin{figure}
\plotone{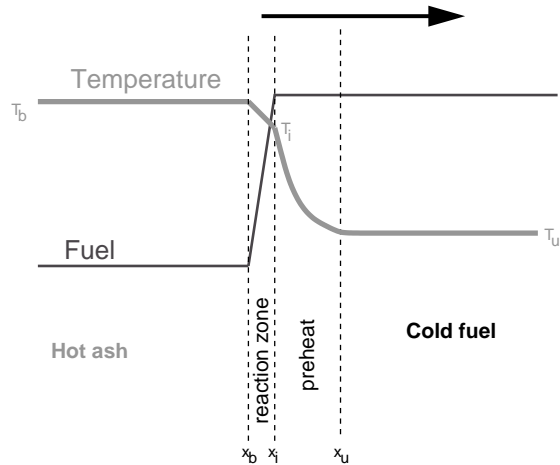}
\caption{\label{fig:flamestruct} 
	  A schematic of an astrophysical thermonuclear flame, propagating
          to the right.  The exothermic reactions are largely concentrated
          in a very thin reaction zone, and energy from that region diffuses
          into the cold fuel over a large preheat zone.
          }
\end{figure}

The structure of an astrophysical flame is sketched in
Fig.~\ref{fig:flamestruct}.  Between the unburned state (which
throughout this paper we will label with a subscript $u$) and the
burned state (subscript $b$), there are two distinct regions.  There is
a `preheat' zone, where energy from the reactions and the already-hot
ash diffuses outwards and heats the incoming (in the frame of the
flame) fuel, and there is a reaction zone where the bulk of the
thermonuclear reactions occur.  The interface between these two zones
--- the state at which burning begins --- we will denote with subscript
$i$.  Because the flame moves very slowly compared to the sound speed in the
domain \citep{timmeswoosley}, the pressure is nearly constant
throughout the unburned fuel and burned ash, $P_b \approx P_u = P$.  It
is possible, however, for the pressure to differ significantly from
this value in the flame structure itself, because of the exothermic
reactions taking place there.

The astrophysical flames of interest to us here are `pre-mixed' ---
that is, the combustible fuel is waiting only for sufficient heat to
ignite.  Thus, the speed of propagation of the flame is determined
completely by two rates --- the rate of energy input by the reaction, and
the rate at which enough heat is diffused outwards to ignite the as-yet
unburned fuel.

Astrophysical premixed flames differ from their terrestrial
counterparts in several respects.  If we were to sketch a terrestrial
flame in Fig.~\ref{fig:flamestruct}, we would have, overlapping with
the thermal diffusion zone, a material diffusion zone where fuel
diffuses inwards towards the fuel-depleted burned zone.  The Lewis
number, $\Lewis$, describes the relative importance of thermal and
species transport across a flame; in terms of the thermal diffusivity
$D_{th}$ and the material diffusivity $D_s$,
$\Lewis = D_{th}/ D_s$.   In the astrophysical
flames we consider here, the Lewis number is of order $10^7$
\citep{timmeswoosley}, whereas in terrestrial flames, the Lewis number
is often of order unity.   Since for astrophysical flames, the effect of
species diffusion is insignificant compared with thermal diffusion,
we neglect species diffusion for astrophysical and some model flames;
in those cases, we informally speak of an infinite Lewis number.
Note, however, that in these numerical simulations, numerical
diffusion will result in some small species diffusion.

Astrophysical flames also have extremely peaked energy-generation rates
--- that is, rates which are strongly dependent on temperature.   This
is usually described by a Zeldovich number,
$\Zeldovich$, defined as
\begin{equation}
\Zeldovich = \frac{T_a}{T_b} \frac{T_b - T_u}{T_b} \enskip .
\label{eq:zeldovichdef}
\end{equation}
Here, $T_a$ is an activation temperature that sets the burning rate by
representing an activation energy for the reaction, such as a potential
energy barrier that must be overcome for the reaction to proceed, and
$T_b$ and $T_u$ are the temperatures of the burned and unburned gas.  In
hydrocarbon-air burning, this number is typically of order 10
\citep{glassman96}; most of the flames we will see here will have
similar ratios, meaning similarly peaked energy generation rates.

Astrophysical flames also typically occur in electron-degenerate
material, meaning that the large amounts of energy input to the flow
(much larger compared to the ambient energy density than in terrestrial
flames) result in comparatively small density changes at the roughly
constant pressure relevant for flame problems.   The ratio of the
burned-gas density to the unburned-gas density will be parameterized by
$\alpha = \rho_u / \rho_b$.  Another factor contributing to a relatively
low $\alpha$ in astrophysical flames is that these flames are powered by
fusion, so that the mean molecular weight of the ash behind the flame is,
unlike many chemical flames, greater than the mean molecular weight of the fuel ahead of it.

The reactions are governed by a strongly-peaked burning rate (ie.,
large $\Zeldovich$), so that in the preheat zone, the fuel can get
quite hot without any significant burning taking place; instead all of
the burning occurs only in the hottest region, in a narrow burning
zone.  In this case, a method for estimating the flame velocity which
dates back to \cite{flamespeed} shows the most important physical
effects.  For a planar, steady flame, one can consider the energy
equation in the reference frame of the flame front itself, and match
boundary conditions between the non-burning preheat zone where thermal
conductivity is important, and a narrow reaction zone.
If one uses a canonical peaked temperature dependence for a burning rate used in
combustion theory, an Arrhenius law (see, for example, textbooks such
as \cite{zeldovich85,glassman96,williams}), which is exponential in the
ratio $-T_a / T$, and consider
our nuclear generation rate to be from a second-order (eg.,
${\mathrm{fuel}} + {\mathrm{fuel}} \rightarrow {\mathrm{products}}$, so
that the reaction depends on the square of the fuel concentration)
then the reaction rate is
\begin{equation}
\enuc = a \rho X_f^2 e^{-T_a/T},
\end{equation}
where $X_f$ is the fuel mass fraction, then one can find
\begin{equation}
S_l^0 = \sqrt{ 2 \frac{\sigma_b}{(C_p)_b^2} \frac{a {X_f}_u^2}{T_a} e^{-T_a/T_b}}.
\end{equation}
where $\sigma$ is the thermal conductivity and $C_p$ is the specific
heat at constant pressure.

This approach is sufficient to get velocities for the astrophysical
flames we describe here to within a factor of a few, and demonstrates
the most important physical effects.   A more careful analysis by
\cite{COpulsations}, using asymptotics and explicitly including an
equation of state and conductivity suitable for completely degenerate
materials, can model the laminar flame speeds to within 25--50\%.

\subsection{Wrinkling}

\begin{figure}
\plotone{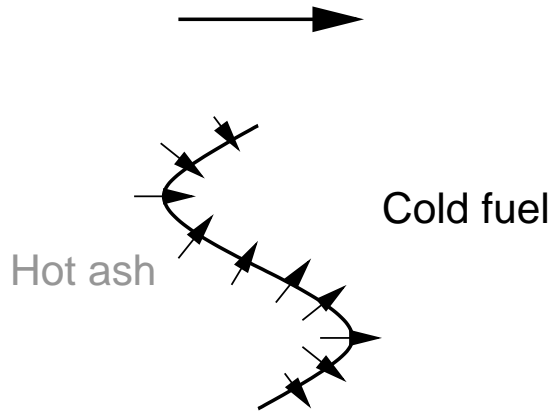}
\caption{\label{fig:heattransport} 
	  A schematic of a wrinkled flame, with the direction of propagation
	  shown by the large arrow.  Heat diffuses from the
	  burning zone to the cool fuel as shown by the little
	  arrows across the flame.  At regions of positive (negative) curvature,
	  heat diffusion is `diluted' (concentrated), and flame propagation will 
          slow down (speed up). }
\end{figure}

Above, we have assumed a planar flame.  Now consider
Fig.~\ref{fig:heattransport}.   In curved regions of the flame, heat
transport into the cold fuel can be concentrated or diluted by simple
geometry of the wrinkling, affecting the flame speed and structure in
those regions.  Since astrophysical flames propagate by thermal
conduction, when the heat transport is concentrated by negative
curvature, we would expect the flame to propagate faster, and when heat
transport is diluted by positive curvature, we would expect the flame
to propagate more slowly.

Note that this figure does not describe cusping of an interface moving
at constant velocity, such as in the Landau-Darrieus instability.
Here, we emphasize that the flame speed itself will change along the
flame; if the flame speed were to vary strongly enough, cusping need
not occur.

The change in geometry can be taken into account by including area
terms into the equations of the previous section, and we can derive
results that include geometric effects due to curvature.   This was
done, for instance, in \cite{lawsung}.   Defining  the strain due to
the curvature with a dimensionless Karlovitz number,
\begin{equation}
\Karlovitz_C = \frac{d A}{A} = \frac{A(x_u) - A(x_b)}{A(x_u)} \approx l_f \frac{1}{A} \frac{dA}{dx},
\label{eq:karlovitzdef}
\end{equation}
where $x_b$ and $x_u$ are the position of the burned and unburned
material, respectively, and $l_f$ is the thickness of the flame.   We
then find to linear order in $\Karlovitz_C$, 
\begin{equation}
S_l = S_l^0 \left ( 1 + \left (\frac{1}{\Lewis} - 1\right) \frac{\Zeldovich}{2} \Karlovitz_C \right ),
\end{equation}
and so for our large-Lewis number flames,
\begin{equation}
S_l = S_l^0 \left ( 1 - \frac{\Zeldovich}{2} \Karlovitz_C \right ).
\label{eq:arrheniuscurvespeed}
\end{equation}

Note that the Karlovitz number is defined so that it is negative if the
area at the burned fluid is greater than that at the unburned fluid,
ie. for regions of negative curvature.   Thus, since
$\Zeldovich > 0$, we get flame speed enhancement in negatively curved
regions, which is as we expect.  Also notice that the effect is
strongest for flames with very peaked temperature dependencies,
where the geometrically-induced difference in temperature diffusion
will have a large effect.

\subsection{Stretch due to flow}

A steady spherical flame will feel the strain described above.
Another contribution to strain in the flame comes from the flow --- for 
instance, a flame propagating outward will also 
be advected outward by the expansion of material
behind it.   Any such flow will also lead to a strain, characterized
by a strain rate $ K = (1/A)(dA/dt)$ where $A$ is the area of
the flame surface.
For the spherical flames we consider here, $A = 4 \pi r^2$, so that
\begin{equation}
K = \frac{2}{r_f} u_{\mathrm{gas}},
\end{equation}
where $r_f$ is the radius of the flame's position, and $u_{\mathrm{gas}}$ is
the flow velocity of the gas at the flame's location.  This strain
can be converted to a dimensionless Karlovitz number by 
considering the strain over a flame crossing time, $l_f/S_l^0$:
\begin{equation}
\Karlovitz_S = \frac{2 l_f}{r} \frac{u_{\mathrm{gas}}}{S_l^0}.
\end{equation}
Note that for a flame in a medium where the flow is zero, this strain
vanishes.

One can also consider a total strain rate, $\Karlovitz_T = \Karlovitz_C
+ \Karlovitz_S$.  We can meaningfully do this for the purposes of
assessing the effect on flame velocity; \cite{groot02} demonstrated
that a single Markstein number can be found that is meaningful for the
combined effects of stretch and curvature, and in fact considering only
the stretch is problematic.  We note here that other works in the
literature choose different sign conventions for $\Karlovitz$ and
$\Markstein$.

\subsection{Effect on Landau-Darrieus growth rate}

The change in flame speed as a function of curvature can 
modify multidimensional flame instabilities.   The Landau-Darrieus
instability \citep{landaulifshitz} is often proposed as a mechanism for
accelerating the burning rate of a flame in a white dwarf.  In 
an otherwise planar flame with a density ratio $\alpha = \rho_u /
\rho_b$, a sinusoidal `wrinkle' of wavenumber $k$ will grow 
in the linear regime with an exponential growth rate
\begin{equation}
\omega =  \frac{\alpha k S_l^0}{\alpha + 1} \left ( \sqrt{\frac{\alpha^2 + \alpha - 1}{\alpha}} - 1 \right).
\end{equation}

One can re-derive the linear theory of Landau-Darrieus growth rates with
the non-constant flame speed given in Equation~\ref{eq:marksteindef}
and arrive at \citep{zeldovich85}
\begin{equation}
\omega = \frac{\alpha k S_l^0}{\alpha + 1} \left ( \sqrt{\frac{\alpha^2 + \alpha - 1}{\alpha} + k l_f \Markstein ( k l_f \Markstein + 2 \alpha )} -1 + k l_f \Markstein \right ),
\label{eq:marksteinld}
\end{equation}
where $l_f$ is the thermal width of the flame.    In fact, some authors
(for example, \cite{clanet98}) have used this to measure the Markstein
number of a flame from an experimentally determined growth rate of
the instability.

From our discussion of the effects of positive and negative Markstein
number, we expect our astrophysical flames to have $\Markstein < 0$.
In this case, the Landau-Darrieus growth rate is reduced, and indeed modes
$k  > (\alpha - 1)/(2 \alpha l_f |\Markstein|)$ are stable against the
Landau-Darrieus instability.   In a negative Markstein number flame,
the flame-curvature relation works against a wrinkle trying to increase
in size; the `peaks' of the flame burn more slowly, because they're at
positive curvature, so they tend to fall back, and the `troughs' of the
flame at negative curvature burn faster and catch back up.

For linear perturbations, the increased burning rate caused by a linear
curvature term at `troughs' in the wrinkle is exactly offset by
a decreased burning rate at the `peaks', so that the total burning rate
remains proportional to the surface area of the flame, as in the constant
flame velocity case.  Thus the total burning rate is decreased by the
decreased growth rate of the instability.  If cusping occurs when the
perturbation becomes nonlinear, then (fast-burning) regions of negative
curvature decrease to almost zero in favor of regions of
(slow-burning) positive curvature, which further decreases the net burning
rate.

Thus, for understanding the behavior of the Landau-Darrieus instability
in astrophysical phenomenon, at least the curvature behavior of the
flame must be modeled.  An alternative is to model the entire flame
behavior, as in \cite{COld}, where a careful analysis of the linear
growth rate and linear instability of a degenerate thermonuclear flame
to the Landau-Darrieus instability was performed.  These authors also
find a cut-off beyond which modes are stable.  Another work,
\cite{niemeyerhillebrandt95}, studied the instability numerically,
although that work obtained anomalous results.  In both analytic cases,
there is a most-unstable mode; there is some evidence
\citep{ldlargescale}  that it is this mode which grows fastest in a
flame going through a complex or turbulent flow field if there is power
on this scale, and indeed there are other suggestions \citep{denet98}
that small scale wrinkling can directly induce larger-scale wrinkling.

\section{NUMERICAL METHODS}
\label{sec:methods}
We consider two classes of flames --- `model' flames and
`astrophysical' flames.  The delineation is reflected in the input
physics.  In the simpler, `model' case, we use a polynomial or simple
exponential expression for the burning and a simple ideal gas equation
of state.  At the complex, astrophysical end, we use a relativistic,
degenerate equation of state (EOS) and a more astrophysically relevant
reaction rate.  In this section we review the physics used in
the astrophysical-flame simulations.   Similar details for the model
flames are described in Appendix~\ref{sec:model-flame-appendix}.

\subsection{Hydrodynamics}

For the hydrodynamics, we use the PPM \citep{ppm} compressible
hydrodynamics module in the \FLASH code \citep{flash}.  PPM is a
widely used Godunov method, which solves the Euler equations in
conservative form using a finite-volume discretization.  Our
implementation of PPM is described in depth in \cite{flash}, and the
PPM module in particular has been tested rigorously in \cite{vandv}.
The Riemann solver is capable of dealing with a general equation of
state, following the method outlined in \citet{colellaglaz}.

To accurately follow reactive flows, a separate advection equation is
solved for each species,
\begin{equation}
\frac{\partial \rho X_i}{\partial t} + \nabla \cdot {\rho X_i v} = \rho R_i \enskip ,
\end{equation}
where $X_i$ is the mass fraction of species $i$, subject to the
constraint that they sum to unity, and $R_i$ is the net rate of change
in the abundance of species $i$ due to the nuclear burning.  $\enuc$
and $R_i$ are related by
\begin{equation}
\enuc = \sum_i {\epsilon_{bi} R_i} \enskip ,
\end{equation}
where $\epsilon_{bi}$ is the binding energy per mass of species $i$.

\subsection{Equation Of State}

Astrophysical flames in degenerate white dwarf material have a complicated
equation of state.  Contributions from the ions, electrons, and radiation
must be accounted for.  The ions are assumed to be fully ionized, and we
use an ideal gas expression for them, as described above.  Finally, the
radiation term is simply blackbody.  Full details of the implementation
of this EOS can be found in \citet{timmesswesty}.

The equation of state is responsible for much of the character of
astrophysical flames.  For the conditions that we consider, the pressure
is dominated by the degenerate electron contribution.  The degeneracy
means that pressure responds only weakly to temperature changes, so the
resulting density jumps behind the flame are smaller than they would be
with a pure ideal gas EOS (and everything else kept the same).

\subsection{Diffusion}

\label{sec:methods:diff}

Diffusion is implemented in \FLASH with explicit time differencing
in an operator-split manner.
The heat flux term is added to the energy evolution equation
\begin{equation}
\label{eq:totenergy}
    \frac{\partial \rho E}{\partial t} 
    + \nabla \cdot \left (\rho E + P \right ) {\bf v}
    = \nabla \cdot (\sigma \nabla T) +
    \rho \enuc{(X_i,\rho, T)}.
\enskip 
\end{equation}
Here, $E$ is the total energy per unit mass, $P$ is the pressure,
${\bf v}$ is the velocity, $T$ is the temperature, $\sigma$ is the
conductivity, $\enuc$ is the energy source term from burning, 
$X_i$ is the mass fraction of species $i$, and $\rho$ is the
total density.  The heat flux $-\sigma \nabla T$ is computed during the
hydrodynamic step and added to the total energy flux computed by the
PPM solver.  The fluxes are then used to update the energy in each
zone.  This implementation preserves the conservative nature of the
algorithm.

Because the diffusion is calculated explicitly, the diffusion term adds
an additional timestep constraint
\begin{equation}
dt \le \frac{1}{2} \frac{\delta x^2}{D_{th}},
\end{equation}
where $D_{th}$ is the diffusion coefficient, $D_{th} = {\sigma}/({\rho C_P})$.
The simulation evolves at either the diffusion timestep or the CFL timestep ---
whichever is smaller. 

For our more realistic thermonuclear flames, the thermal diffusion is
the only diffusive process modelled.  For some of our model flames,
we allow finite $\Lewis$, and so species diffusion is also calculated;
the computation proceeds in the same manner.  This is described in
Appendix~\ref{sec:model-flame-appendix}.

Since these diffusion modules were not included in the original \FLASH
code paper \citep{flash}, we include in Appendix~\ref{sec:appendix}
tests of the correctness of the diffusion operators in Cartesian and
spherical coordinates.

\subsection{Conductivity}

For testing purposes and for our model flame propagation problems
in Appendix~\ref{sec:model-flame-appendix}, we use a simple constant
diffusivity to describe the heat transfer.  However, the conductivity
for astrophysical flames of interest to the Type Ia problem is much more
complicated.  For degenerate carbon, electron-electron and electron-ion
collisions are important processes in heat transport.  The conductivity in
the post-flame state can be some 3 orders of magnitude higher than that of
the pre-flame state, making a constant conductivity a bad approximation.
We use a stellar conductivity routine that includes these processes as
well as radiative opacities \citep{timmes_conductivity}.

\subsection{Burning}

We experiment with several reaction rates and networks for our flame
simulations.  In this paper, we use one-step irreversible reactions,
with increasingly complex reaction rates.  In all cases the energy
release from the nuclear burning is computed in an operator split
fashion.  After the hydrodynamics are evolved, the reaction network
computes the energy release and change in nuclear abundances over the
course of the hydro timestep.  This energy release is then added to the
internal energy of that zone, and a new temperature is computed.
For these fully-resolved flames, the CFL-limited timestep from the
explicit hydrodynamics is much smaller than the timescale to
significantly change the temperature by burning, so that explicit
coupling is not needed.   We describe the burning rates for our
astrophysical flames here; model flame burning is described in 
Appendix~\ref{sec:model-flame-appendix}.

\subsubsection{One-step Carbon Burning}
\label{sec:methods:burning:cf88}

The astrophysically-motivated reaction we use is a one-step $^{12}$C +
$^{12}$C reaction.  The rate comes from \citet{cf88} (we refer to this
as the CF88 reaction network) and converts two $^{12}$C nuclei into a
single $^{24}$Mg nuclei, releasing $5.57\times 10^{17} \mathrm{erg} \ \mathrm{g}^{-1}$.   Following the notation of \citet{cf88},
the reaction rate is
\begin{equation}
\frac{\mathrm{d}X_C}{\mathrm{d}t} = -\frac{1}{12} X_C^2 \rho N_A\lambda ,
\end{equation}
where 
\begin{equation}
N_A \lambda = 4.27 \times 10^{26} \frac{T_{9,a}^{5/6}}{T_9^{3/2}}
      \exp\left \{ {\frac{-84.165}{T_{9,a}^{1/3}} - 2.12 \times 10^{-3} T_9^3} \right \},
\end{equation}
where $T_9 = T/10^9 \ \mathrm{K}$, $T_{9,a} = {T_9}/{1.0 + 0.0396 T_9}$,
and $X_C$ is the mass fraction of $^{12}$C.
For each $^{24}$Mg nucleus created, 13.933 MeV is released.

This is the main reaction involved in a pure carbon flame in a white
dwarf.  We treat all other species (eg. $^{16}$O) simply as inert
dilutants that serve only to reduce the effective burning rate.  We
assume here that other reactions (eg.
$^{24}$Mg($\alpha$,$\gamma$)$^{28}$Si burning), if modeled, would occur
well behind the flame and would not be as important in setting the
properties of the flame.

With this reaction mechanism, we neglect any ion screening, for
simplicity.  Screening is not an important effect except at low
temperatures or high densities.  At high densities, a one-stage
reaction as modeled by this rate might have other problems; late stage
reactions could be important.   Conversely, low temperatures are not of
interest to us for these flame-propagation problems.   Thus we
restrict ourselves to regimes where screening is unimportant.

%
%

\subsection{Boundary Conditions}

Boundary conditions for subsonic flows in a compressible code are
non-trivial.  The zero-gradient/outflow condition usually used is
insufficient, since not all the characteristics at the edge of the
computational domain will be leaving the domain.  This means that some
information can enter the domain from the boundary and introduce noise
into the flame solution.

Boundary conditions for a subsonic outflow have been proposed
(see for example \citealt{bc1,bc2}), but no well-accepted general solutions
exist.  These boundary conditions try to extend the solution
into the boundary conditions in such a way that information is not
transmitted into the domain from the boundaries.  We choose
problem-specific boundary conditions that allow most waves to leave the
domain.  We first use the fact that, to a good approximation, the
pressure is constant in the burned and unburned regions, and since the flames
are not in a closed vessel, the pressures will be constant in time.  This allows
us to set the pressure in the boundary conditions as a constant.
Furthermore, since we want heat to diffuse out, the temperature 
boundary condition is simply set to be zero-gradient.  The 
boundary conditions for the mass fractions are also 
zero-gradient.  This gives us enough information to find a 
thermodynamically consistent
density through our equation of state.  Finally, the velocity in the
boundary conditions is found by enforcing mass conservation,
\begin{equation} 
\rho_i v_i A_i = \rho_{i+1} v_{i+1} A_{i+1} 
\end{equation} 
if the velocities are out of the domain, or zero if the velocities are
inwards.  Here $A_i$ is the area of the face through which the material
flows;  in planar geometry, this cancels out, but it is important in
the spherical geometries we consider.  A demonstration that this
boundary condition functions in the way we'd like is given in
Appendix~\ref{sec:appendix}.

For our spherical runs, we use at $r = 0$ a reflecting boundary condition.
This boundary condition is required by the geometry; spherical symmetry
requires, for any quantity $f(r)$ other than the radial velocity, $f(r)
= f(-r)$, and for the radial velocity, $v_r(r) = -v_r(-r)$.  This is
exactly the reflecting boundary condition as implemented in the \FLASH{}
code \citep{flash}.

\subsection{Adaptive Mesh Refinement}

\FLASH employs an adaptive mesh through the PARAMESH package \citep{paramesh}, 
allowing it to put computational zones where the resolution is needed
and to follow smooth flow with less resolution.  The \FLASH mesh
adapts by dividing the domain in half in each dimension, creating new
blocks (2 in 1-d, 4 in 2-d, 8 in 3-d) that are logically the same as
their parent (typically containing 8 computational zones in each
direction) but with twice the spatial resolution.  Each new block is
checked to see if more refinement is needed, and if so, it is further
subdivided.  When a block is refined, the newly created children need
to be initialized with information from their parent.  To obtain
higher-order accuracy, a quadratic polynomial is fit to the data of
the parent zone and the two zones on either side of it.  This
procedure is described in detail in Appendix~\ref{sec:appendixb}.

When evolving a flame, it is only necessary to put the resolution near
the reaction and diffusion zones.  We refine when the second
derivatives of density or pressure and are large (the former is
expected to be large at the fuel/ash interface, and the latter in the
burning region).  We also refine on the second derivatives of the
nuclear energy generation rate and temperature.  The methods for
evaluating the derivatives and determining their magnitude is
discussed elsewhere \citep{flash}.  These criteria ensure that the
region around the flame is highly resolved.

\section{EXPERIMENTS}
\label{sec:experiments}
\subsection{Overview}

To numerically investigate the effect of curvature on astrophysical
flame speed, we conduct numerical experiments of 1-d model and
astrophysical flames running out from, and in toward, the center of a
spherical domain, as has been done in the terrestrial combustion
community experimentally and computationally  (eg., 
\citealt{karpov97,muller97,kwon01,
aung98,gu00,hassan98,bradley96,bradley98,sun99}).  Doing so allows us
to examine the flame speed and structure for a range of curvatures and
strain rates, both positive and negative, by taking snapshots of the
flame at varying positions (and thus curvatures) throughout the domain.
We simulate the same flames in planar geometry for comparison.

Focusing on 1-d is computationally more efficient and allows us to
separate the pure strain and curvature effects from more complicated
multidimensional instabilities, which will be the focus of a later
paper.  In addition, for the present work, we studied the effects of
resolution on the experiments and measurements. The details are discussed
in Appendix~\ref{sec:resstudy}.  In the numerical experiments we present,
the flames were sufficiently well-resolved to accurately model the flame
structure and speed.

We begin with the simpler model flames, adjusting parameters to ensure
we understand the effects of each.   For each set of input physics
(EOS, reaction mechanism, and model parameters), we simulate three
flames --- a planar flame, a spherical flame propagating out from the
center, and a spherical flame propagating toward the center.   We then
continue on with simulations of a simplified astrophysical reaction
rate with a complex EOS.

In each case, we ignite the flame by placing hot ash --- at a
temperature corresponding to the ambient planar burned temperature of
the flame, unless stated otherwise --- alongside the cold fuel we wish
to burn, in pressure equilibrium.  Neglecting to place the fuel and ash
in pressure equilibrium will generate strong pressure waves, especially
in non- or partially degenerate fluids, giving spurious results --- one
is no longer studying a flame, but a transient pressure-driven reaction
front.  We believe this is the reason for the anomalous results found
in \cite{niemeyerhillebrandt95}, which have not been reproduced in work
since.

For planar flames, and for the outer boundary conditions for our flames in
spherical coordinates, we use the constant-pressure boundary conditions
described in \S\ref{sec:methods}.  For the inner boundary conditions
in spherical coordinates, we use reflecting (or `symmetry') boundary
conditions at $x=0$ to be consistent with the spherical geometry.
These boundary conditions are consistent with those used in the
terrestrial combustion literature, {eg., \citep{sun99}.

\subsection{Measuring flame position}

One of the most fundamental things we need to do is to measure 
the flame's position ($r_f$) at different times.  Since flames have finite
thickness, `the' flame position is ambiguous.  Following
\cite{groot02}, we take the flame's position to be that of the reaction
zone, which is less sensitive to strain and curvature effects than, for
instance, the outside of the preheat zone.  Then a consistent Markstein
number can be calculated that includes the effects of curvature and
stretch.   Since even the reaction zone has finite thickness, we
consider several measures for the position of the reaction zone.   The
first is the most obvious, the position of maximum $\enuc$.  This has
the advantage of simplicity, but could potentially be noisy, since
extrema are generally not numerically well behaved.  Another is $\avg{\enuc
x}/\avg{\enuc}$, an $\enuc$-weighted position.  Since $\enuc$ is very
strongly peaked, this should localize the flame.  Clearly, in the limit
of a delta-function $\enuc$, this measure reduces to finding the $x$ of
maximum $\enuc$, but as an integral quantity, it will vary more smoothly.
A third measure is where the fuel concentration is most
rapidly changing --- the location of maximum $d X_f / d x$.  A fourth
is a temperature criterion, the location of maximum $d^2 T/dx^2$.  In
Fig.~\ref{fig:position-measures}, we show the results of using these
measures for a simulation with a KPP flame and an Arrhenius flame.  The
KPP flame has a reaction zone approximately $6~ {\mathrm{cm}}$ wide in
a domain of $500~{\mathrm{cm}}$, and the Arrhenius flame has a reaction
zone of approximately $1/2~{\mathrm{cm}}$ in a domain of
$100~{\mathrm{cm}}$.

\begin{figure}
\plotone{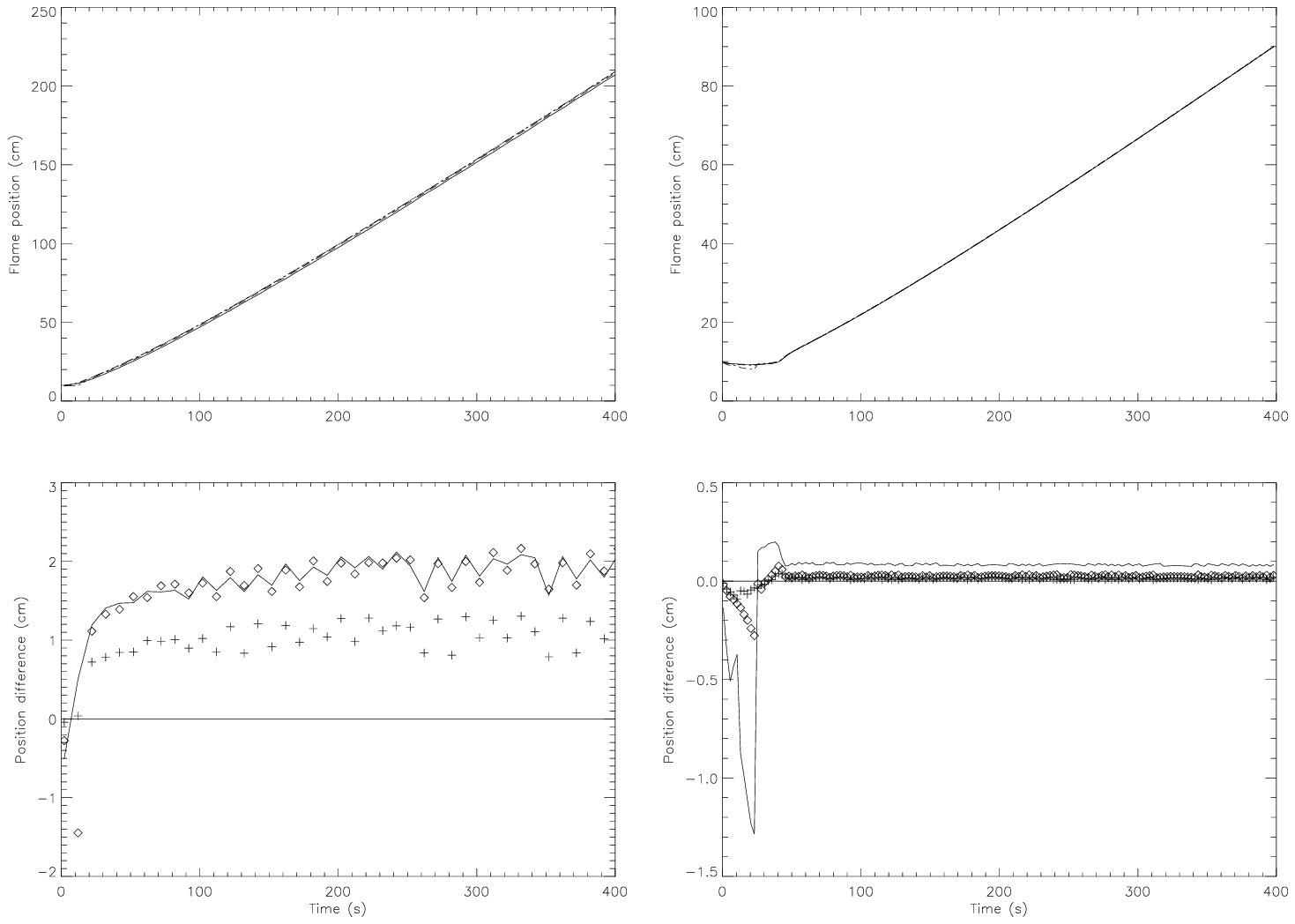}
\caption{\label{fig:position-measures} Position versus time, 
         measured four different ways, for both a KPP (left) 
         and an Arrhenius (right) flame.   Both flames have $\Lewis = \infty$
         and are propagating spherically outwards.  Results from all four
         position measures are plotted on the top.  On the
         bottom are plotted differences from using $\avg{\enuc x}/\avg{\enuc}$.
         The three other methods are: maximum $\enuc$ (+), 
         maximum $d^2T/dx^2$, (-), and maximum $dX_f/dx$ ($\diamond$).
         The KPP flame has a reaction zone of $6 {\ \mathrm{cm}}$,
         and the Arrhenius flame has a reaction zone of $1/2 {\ \mathrm{cm}}$.
         The Arrhenius reaction takes about 50 seconds to ignite a 
         self-sustaining flame; before then, any measure of `flame position' 
         is necessarily meaningless.
    }
\end{figure}

Because the KPP reaction zone is so wide and includes the preheat
zone, finding an unambiguous `location' is especially difficult.  The reaction
rate depends as strongly on fuel as on temperature, so the fuel-change
rate is fastest right at the front of the flame.  Also, the energy
generation rate is large in a wide region, so averaging over $\enuc$
produces a position that lags well behind the front of the zone.

For the Arrhenius rate, however, once ignition occurs, all of
the direct burning measures ($\enuc$, $\avg{x \enuc}/\avg{\enuc}$, and
$dX_f/dx$) fall almost exactly on top of each other, with only
the temperature measure leading slightly, as the temperature must
start diffusing outward from the burning region to change the
temperature's second derivative.    

In light of the above, we use the $\enuc$-weighted position as a marker
for the position of the reaction zone, and thus the flame, for flames
with strongly-peaked reaction networks (Arrhenius, CF88) as it
generates fairly clean velocities through differencing and is
otherwise essentially degenerate with the very intuitive
maximum-of-$\enuc$ measure of position.   For KPP, we use the position
of maximum $\enuc$.

\subsection{Measuring flame thickness}

The thermal thickness of the flame is an important quantity, setting
the length scale for relevant microphysical effects.  Since this width
is set by a diffusive process, however, its precise start and end are
poorly defined.  We use two measures of the flame thickness, both used
in the literature, that span the range of reasonable measures of the
thickness.   Both are based on the change in temperature from the
burned to unburned state, $\Delta T = T_b - T_u$.

The first way we use to measure flame thickness, method I, is the
`10/90' approach used for instance in \cite{timmeswoosley}.  It
measures the distance from where the temperature crosses 10\% above the
unburned temperature to 90\% of the way to the burned temperature
\begin{equation}
l_f^{(I)} \equiv x(T=T_b - .1 \Delta T) - x(T = T_u + .1 \Delta T).
\label{eq:methodonethickness}
\end{equation}
This gives quite wide thermal widths, as it
measures well into the diffusive tails of the thermal structure.

The second measure, method II, used for instance in \cite{sun99},
finds the width by dividing the
temperature difference by the maximum temperature gradient in the
flame
\begin{equation}
l_f^{(II)} \equiv \frac{\Delta T}{\max |\nabla T|}.
\end{equation}
This measure gives the thinnest meaningful thermal thickness of the flame,
as any smaller thickness must imply a larger thermal gradient
than any that exists in the flame structure.    
We report both flame thicknesses and use them to constrain measurement
uncertainties in derived quantities that depend on flame thickness.

\subsection{Measuring flame velocity}
\label{subsec:flamevel}

The velocity we are interested in is not the change of flame position
over time ($d r_f / d t$), but the rate at which the flame is consuming
fuel.   This means we want to measure the flame speed with respect to
the fuel. The expansion of hot burned ash smoothly generates a velocity
field as the flame passes through.  In the case of spherical geometry,
the velocity of the material between the flame and the origin is zero,
since the material has no place to go, and thus the only nonzero
velocities are exterior to the flame.   In the open planar geometry,
material can flow freely through both boundaries.   In both cases, the
continuity condition requires $A \rho u = {\mathrm{const}}$ on either
side of the flame.  Fig.~\ref{fig:gasvelocity} sketches idealized
velocity profiles for the three configurations.

\begin{figure}
\plotone{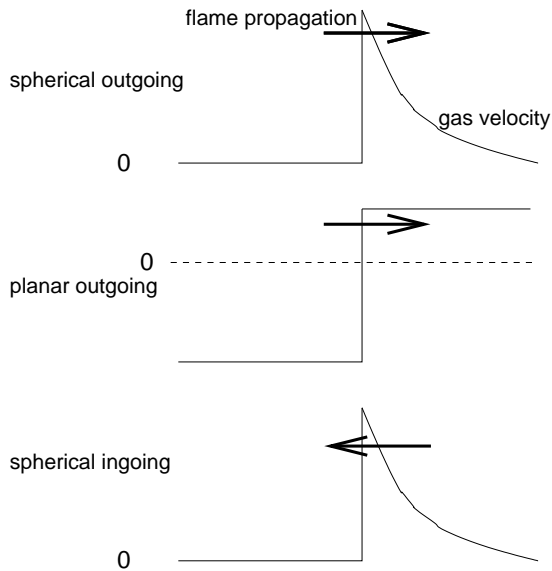}
\caption{\label{fig:gasvelocity} 
         The gas velocity structure of a propagating flame 
         with the flame propagating outward through a spherical
         domain (top), through a planar domain (middle), and inward in
         a spherical domain (bottom).   In the closed geometries, the
         left state, either burned or unburned, has a velocity of zero
         imposed by the boundary and the continuity condition, 
         $A \rho v = {\mathrm{const}}$.
        }
\end{figure}

To find the propagation velocity into the fuel, one approach is to find the
change in the flame's position over time (by differencing, or by fitting
and then differentiating) and to subtract off the gas velocity (which can
be directly measured).  This can be problematic, however, especially
in the outward-expanding flame case.   The expanding gas, especially
for large density contrasts, will be a large component of $d r_f/
d t$, and thus any subtraction is likely to be noisy ---  especially since
the flame velocity and the gas velocity are found in different ways.
The situation is worse in the spherical case, as gas velocity has a
non-constant spatial structure, adding difficulty to measuring `the'
gas velocity.

Another approach, used in the chemical literature (eg., 
\citealt{aung97,hassan98}), is to note that if the flame is a discontinuity,
it must separate two regions with the same
momentum flux in the frame of the flame.   Given that in the lab frame, the velocity interior to
the flame is zero, the flame propagation velocity must be $(\rho_b /
\rho_u) (d r_f / dt)$.   However, this is an approximation that assumes
an infinitely thin flame, and we are measuring effects explicitly
due to the finite structure of the flame.   A better approximation, $(A_b
\rho_b) / (A_u \rho_u) ( d r_f / d t )$, includes the effects of geometry
but requires locating `the' position of the unburned and burned states,
adding uncertainty due to the ambiguity of those locations.

Since what we are really interested in is the burning rate, which 
is directly measurable, another approach is simply to measure the
total burning rate, $\dot E = \int \enuc \rho dV$.  If one then
assumes that the reaction region is thin (a better approximation than
assuming the thermal structure of the flame to be thin), then one can
compute a per-area burning rate, $\dot E / (4 \pi r_f^2)$.  Knowing the
per-area burning rate and the chemical energy per volume ($\Delta e
\rho X_f / \mu_f$) of the unburned state, where $\Delta e$ is the
binding energy per fuel particle and $\mu_f$ is its mass, one can find
the speed at which the flame must be propagating through the unburned state to
generate the measured energy, $\dot E \mu_f / (4 \pi r_f^2 \rho X_f
\Delta e)$.

A comparison of several methods of measuring velocity is given in
Fig.~\ref{fig:velmeasures}.  These plots were made for a typical flame with 
typically noisy
velocity data, (the $\Lewis = 5$, $\Zeldovich = 5$ outwardly propagating
Arrhenius flame) showing the flame's propagation velocity
through the fuel as a function of time; ignition occurs at $t \approx
50\ {\mathrm{sec}}$.   To calculate the area-corrected expression
$(A_b \rho_b) / (A_u \rho_u) ( d r_f / d t )$, we (rather crudely)
assumed $A_b \approx 4 \pi (r_f - l_f^{0(I)}/2)^2$ and $A_u \approx 4
\pi (r_f + l_f^{0(I)}/2)^2$.  Note that ignoring the thickness of the
flame spuriously increases the measured effect of curvature in the
$(\rho_b /\rho_u) (d r_f / dt)$ velocity measure.

\begin{figure}
\plotone{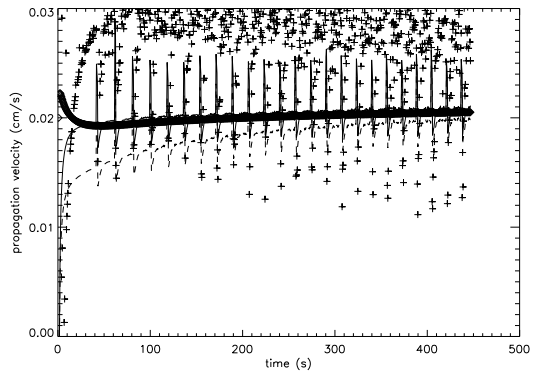}
\caption{\label{fig:velmeasures} 
         The propagation velocity -- the speed of the flame into the fuel --
         of a $\Lewis = 5$, $\Zeldovich \approx 5$ Arrhenius flame propagating
         outwards in a spherical domain.   Shown is $(d r_f / d t) (\rho_b/\rho_u)$
         (`$-\,-$'), $(A_b \rho_b)/(A_u \rho_u) (d r_f / d t)$ (`---'),
         velocity from total $\enuc$ (`$\diamond$'), and $(d r_f / d t) - v_{\mathrm{gas}}$
         (`+').
        }
\end{figure}

Another thing to note in Fig.~\ref{fig:velmeasures} is that in the
direct velocity measure $d r_f/dt$, one can see pulsations due to an
essentially translational 1-d instability in a flame with large Lewis
number (see, for instance, \cite{zeldovich85,COpulsations}).  While
this slightly affects the instantaneous position of the flame front,
and, therefore, velocity measures based on differencing the
flame position, it has little effect on the total
burning occurring in the flame, and thus in velocity measures based on
bulk burning rate.

\subsection{Igniting the flame}

Our flames are ignited by placing a hot region of ash next to the fuel,
and letting the heat diffuse into the fuel until the fuel ignites a
propagating wave.   To ensure that the details of the ignition process
do not affect the later flame propagation (for instance, by inducing
significant pressure waves, or by `overdriving' the flame), we ran
simple tests of using different temperatures to ignite the
flame.   Results for a $\Lewis = \infty$ and $\Lewis = 1$ Arrhenius
flame are shown in Fig.~\ref{fig:ignitioneffects}, where we vary the
ignition temperature by a factor of four.   The two flames are shown because
they have distinct ignition mechanisms --- the $\Lewis = \infty$ flame
ignites due to thermal diffusion, eventually heating a fuel layer to
the point where significant burning occurs, whereas the $\Lewis = 1$
flame ignites due to combustible fuel diffusing into the already hot
ash and igniting.  We see that changing the
ignition temperature affects the time-to-ignition for these flames,
but it does not affect later flame propagation.

\begin{figure}
\plottwo{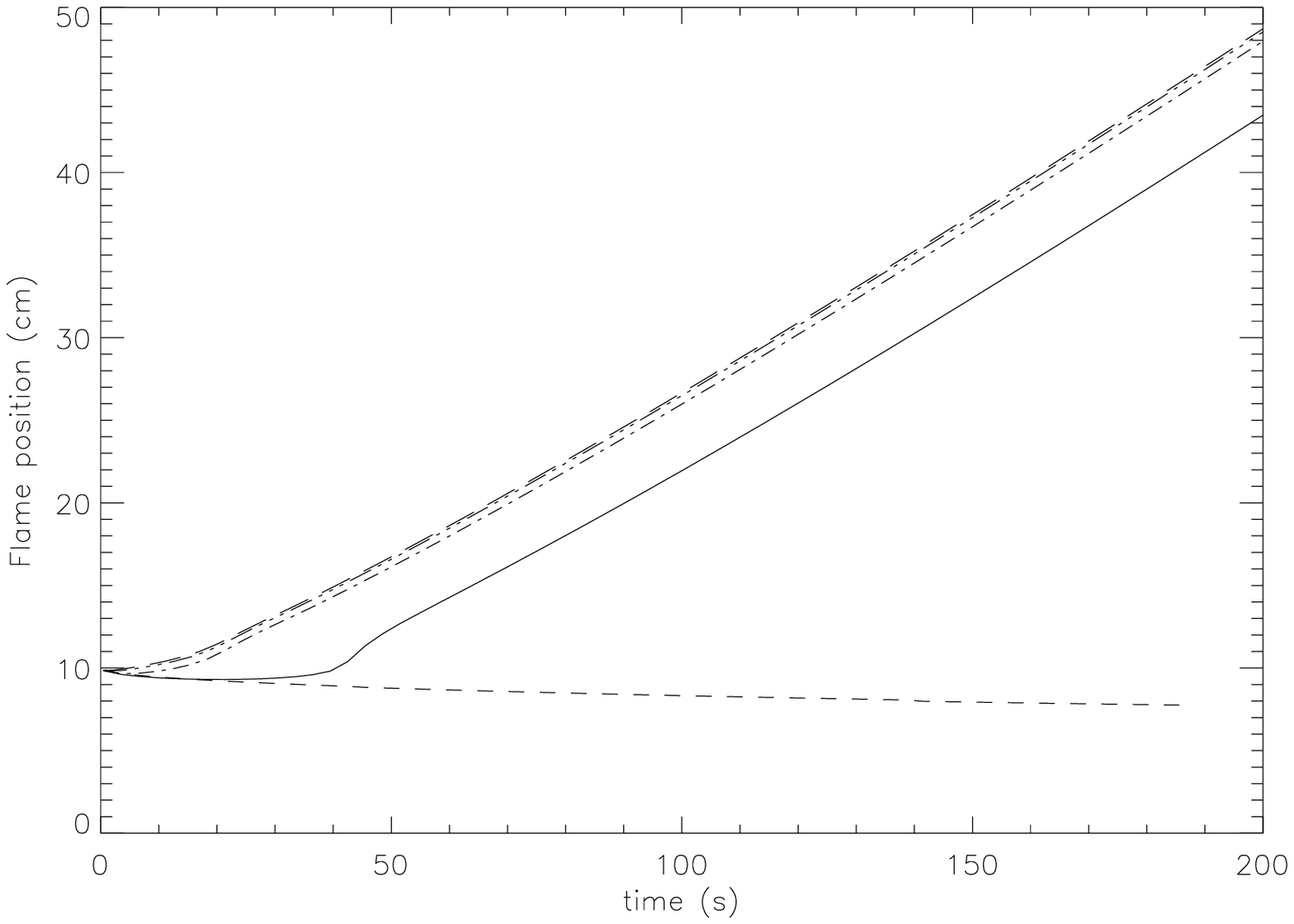}{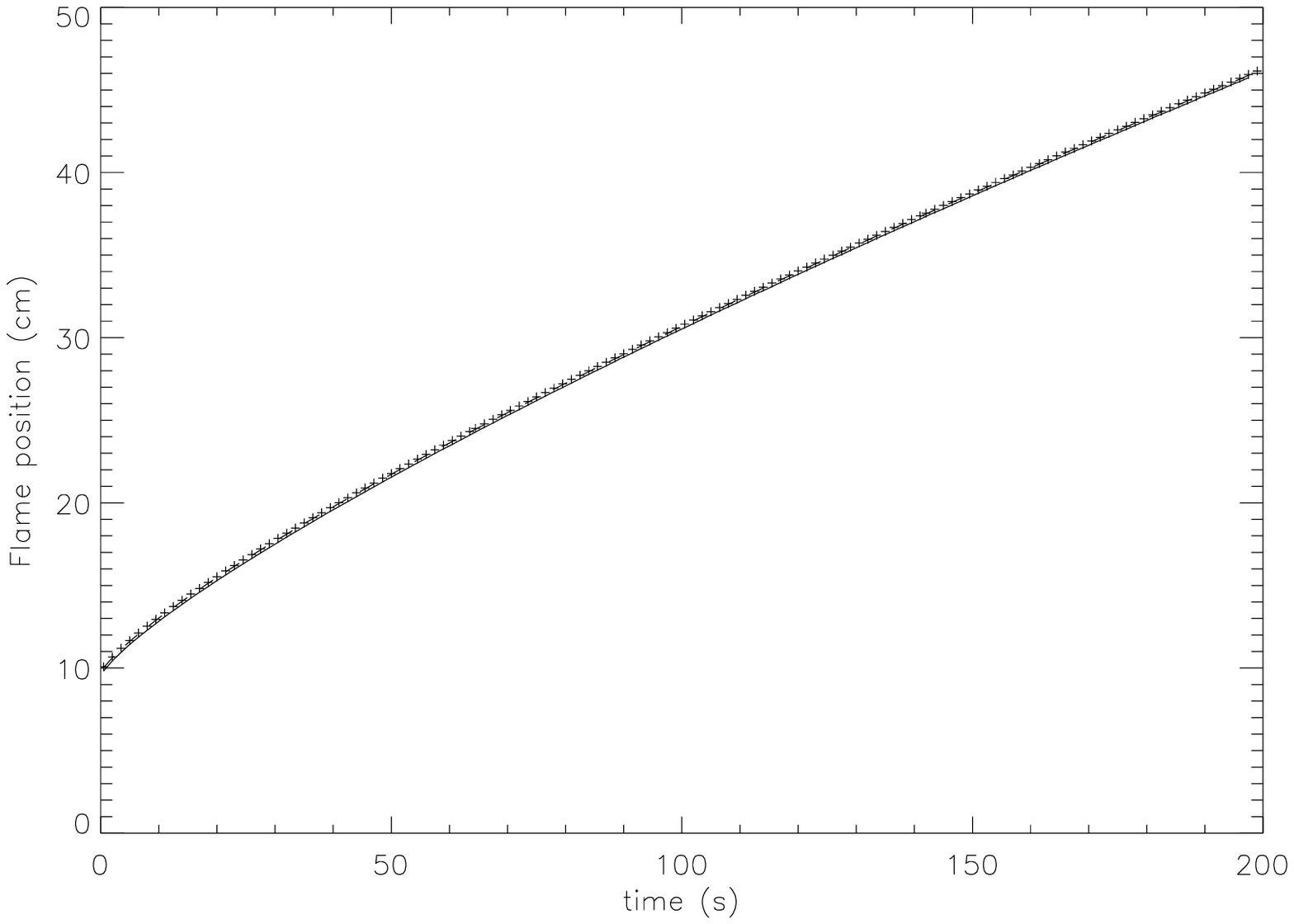}
\caption{\label{fig:ignitioneffects} Position versus time,
	 for a spherically-expanding Arrhenius flame with $\Lewis =
	 \infty$ (left) and $\Lewis = 1$ (right).	  Ignition for
	 the two Lewis numbers proceeds differently.  For each flame,
	 the region $x \in [0,10]$ was hot ash at temperatures (bottom
	 to top) $T=0.64, 0.96, 1.28, 1.92, 2.56$, (only $0.64$, $0.96$, and
	 $1.28$ for the $\Lewis = 1$ flame) with densities set to attain 
	 pressure equilibrium with the cold fuel at $\rho = 1.0, T = 0.1$
	 in the rest of the $100 \ \mathrm{cm}$ domain.	The adiabatic
	 flame temperature is $T_b = 0.64$.  As can be seen, the ignition
	 times can vary significantly in the $\Lewis = \infty$ case (with
	 the lowest-temperature case never igniting a propagating flame
	 at all), but once ignition has occurred, the flame propagates
	 with the same velocity.   For the $\Lewis = 1$ flame, even the
	 ignition time is nearly constant.
} 
\end{figure}

\subsection{Measuring curvature strain rate and Markstein number}

As discussed in \S\ref{sec:theory}, we are interested in the
total dimensionless strain, which for our two spherical flames are
\begin{equation}
\Karlovitz_T  =  \Karlovitz_S + \Karlovitz_C 
              =  \frac{2 l_f}{r_f}\frac{u_{\mathrm{gas}}}{S_l^0} + \frac{2 l_f}{r_f}
              =  l_f \frac{2}{r_f} \frac{d r_f/dt}{S_l^0}
\end{equation}
since $d r_f/d t = u_{\mathrm{gas}} + S_l^0$.  We have described
measuring $r_f$ above, and $d r_f/dt$ can be calculated by
differencing as long as care is taken to have sufficiently
time-sampled data.  $S_l^0$ can be measured from our planar flame
simulations after we have settled into a steady state.    We
have described methods to measure $l_f$, but instead we choose to
simply calculate $\Karlovitz_T / l_f$, leaving
\begin{eqnarray}
S_l & = & S_l^0 \left ( 1 + \left( l_f \Markstein \right) \left (\Karlovitz_T/l_f\right ) \right ),  \\ 
    & = & S_l^0 \left ( 1 + \MarksteinLength \Karlovitz_T/l_f \right) 
\end{eqnarray}
so that a physical dimensional quantity, the Markstein length
($\MarksteinLength = l_f \Markstein$) can be measured directly by
fitting, and then a dimensionless Markstein number can be calculated
using any chosen thermal width of the flame.

\section{RESULTS}
\label{sec:results}
The goal of this study is to investigate the effect of curvature on the
speed and structure of several types of flames. As described above, the
effects define the Markstein relation, so the majority of our results
consists of quantifying this relation for each of the flames we
consider.  Best fits were calculated using a maximum likelihood linear
least squares fit, and errors quoted in that fit come from variances of
the fit parameters unless stated otherwise (see for instance
\citealt{datareduction}).    Astrophysical flame results are
presented here; model flame results can be found in 
Appendix~\ref{sec:model-flame-appendix}.

The planar flames were run first to determine the flame width and
speed accurately.  Experimentation showed that we need at least 10
computational zones in the 10/90 thermal width.  We ran all of these
flames with about 20 points in this thermal width (as determined
from the planar flame runs).  Several densities were run, both in
pure carbon and half carbon/half oxygen, and the flames were run both
inward and outward in radius.   Data on the planar flames is given in
Table~\ref{table:cf88dataother}.

\begin{figure}
\plotone{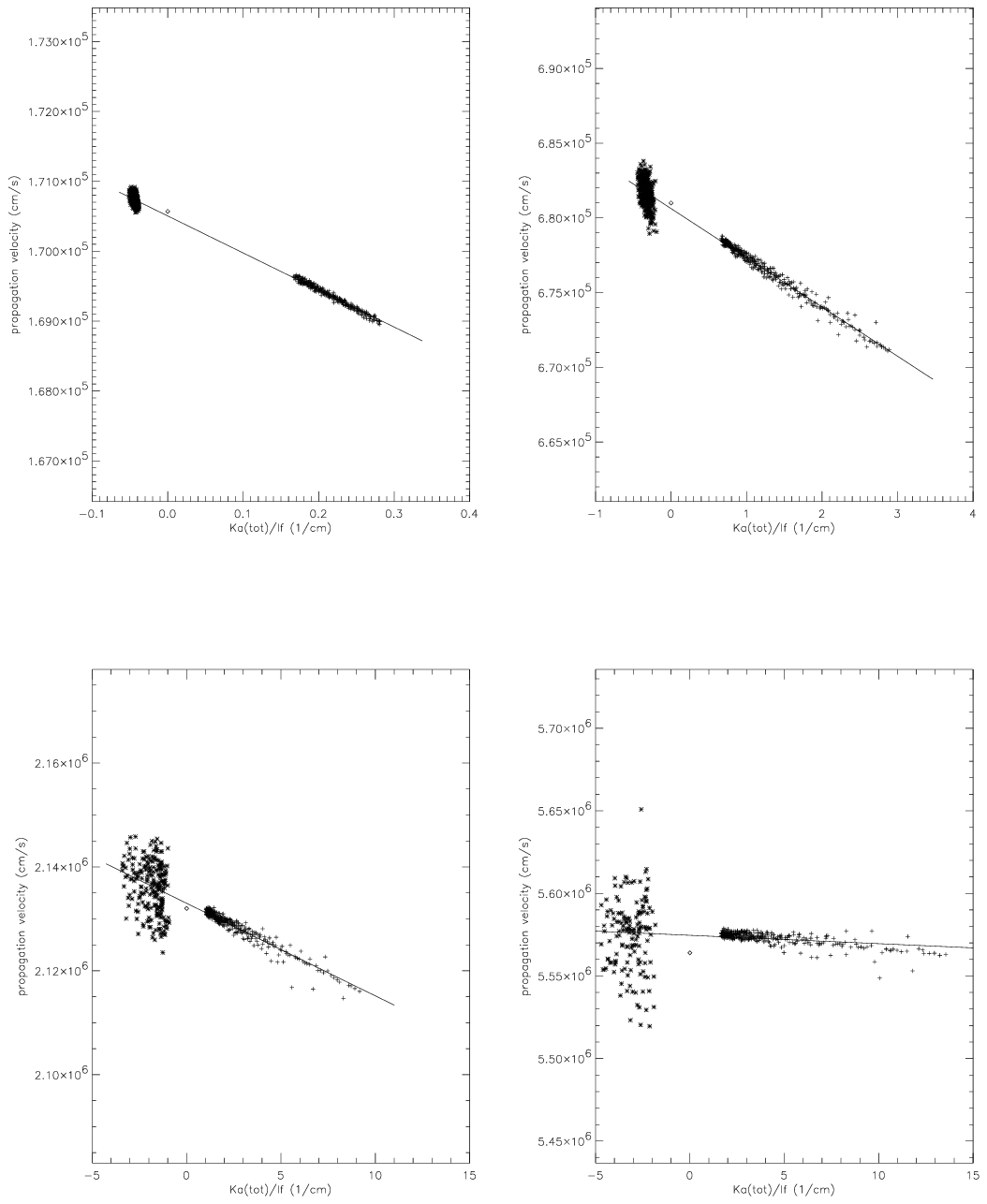}

\caption{\label{fig:cf88flamecarbon} Flame speed versus
         $\Karlovitz_T / l_f$ for a $2.5\times 10^7 \gcc$ (top left), $5\times 10^7 \gcc$ (top right),
    $1\times 10^8\gcc$ (bottom left), and $2\times 10^8\gcc$ (bottom right)  pure carbon flame.  }
\end{figure}

\begin{figure}
\plotone{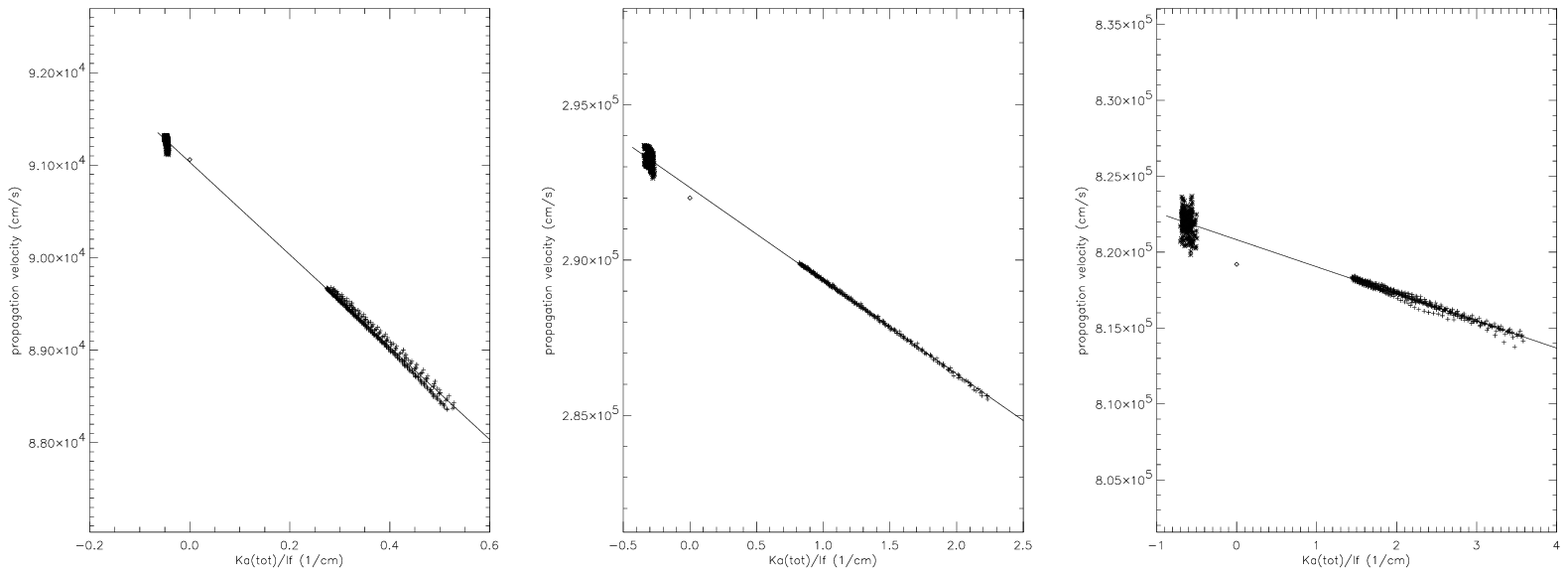}

\caption{\label{fig:cf88flamecarbonoxygen} 
         Flame speed versus $\Karlovitz_T / l_f$ for a
         $5\times 10^7 \gcc$ (left), $1\times 10^8 \gcc$ (center) and
         $2\times 10^8\gcc$ (right) 0.5 C/ 0.5 O flames.  }
\end{figure}

\begin{table}[H]
\begin{center}

\begin{tabular}{rrddrrrr}
\tableline
\tableline
\multicolumn{2}{c}{Unburned state} & \multicolumn{2}{c}{} & \multicolumn{2}{c}{Unburned} & \multicolumn{2}{c}{Burned} \\
\multicolumn{1}{c}{$\rho$}& $X_{\mathrm{{}^{12}C}}$& \multicolumn{1}{c}{$\Zeldovich$} 
                     & \multicolumn{1}{c}{$\alpha$}
                     & \multicolumn{1}{c}{$P$ (dyn cm${}^{-2}$)} 
                     & \multicolumn{1}{c}{$T$ (K)} 
                     & \multicolumn{1}{c}{$P$ (dyn cm${}^{-2}$)} 
                     & \multicolumn{1}{c}{$T$ (K)} \\ 

\tableline
$2.5 \times 10^7$ & 1 &  13.4 & 2.17 & $3.14 \times 10^{24}$ & $1.0 \times 10^7$
                                   & $3.14 \times 10^{24}$ & $4.05 \times 10^9$ \\
\tableline
$5 \times 10^7$ & 1/2 & 13.5 & 1.47 & $8.27 \times 10^{24}$   & $1.0 \times 10^7$
                                    & $8.27 \times 10^{24}$  & $3.89 \times 10^9$\\

                & 1   & 12.4 & 1.89 & $8.27 \times 10^{24}$ & $1.0 \times 10^7$
                                    & $8.27 \times 10^{24}$ & $4.86 \times 10^9$\\
\tableline
$1 \times 10^8$ & $1/2$ & 12.8 & 1.36 & $2.15 \times 10^{25}$  & $1.0\times 10^7$
                                      & $2.15 \times 10^{25}$  & $4.49 \times 10^9$\\

                & 1   & 11.1 & 1.69 & $2.15 \times 10^{25} $  & $1.0\times 10^7$
                                    & $2.15 \times 10^{25} $ & $5.79 \times 10^9$\\
\tableline
$2 \times 10^8$ & 1/2 & 11.9 & 1.29 & $5.54 \times 10^{25}$  & $1.0 \times 10^7$
                                    & $5.54 \times 10^{25}$ & $5.16 \times 10^9$\\

                & 1   & 9.78 & 1.54 & $5.54 \times 10^{25}$  & $1.0 \times 10^7$
                                    & $5.54 \times 10^{25}$  & $6.82 \times 10^9$ \\

\tableline

\end{tabular}

\end{center}
\caption{Flame properties of the planar CF88 flames}
\label{table:cf88dataother} 
\end{table}

Fig.~\ref{fig:cf88flamecarbon} shows the flame velocity vs. scaled
dimensionless strain for flame in a pure carbon medium, and
Fig.~\ref{fig:cf88flamecarbonoxygen} show results for a 50/50
Carbon/Oxygen medium.  The results are summarized in
Table~\ref{table:cf88flamedata}.  Quoted errors on the Markstein
length come from uncertainties in the fit.  As is the case with the
model flames, and is described in Appendix~\ref{sec:model-flame-appendix},
the data for the inward propagating flames is far more
noisy than the outward propagating flames due to pressure waves
from ignition transients being trapped inside the unburned region.
The noise is, however, reasonably symmetrically distributed about the
line fit.  We do not need the inward data to get a Markstein length,
as the outward data alone is enough to provide this, so we can do
fits of just the outward data to see how it compares to the complete
data set.  In the most severe cases, the difference in the Markstein
length computed with and without the inward propagating data is 10\%.
For the $1\times 10^8\gcc$ pure carbon flame, $\MarksteinLength = -(8.366
\pm .238) \times 10^{-4} \ \mathrm{cm}$ when all of the data is used,
and $\MarksteinLength = -(9.003 \pm .12) \times 10^{-4} \ \mathrm{cm}$
when only the outward propagating data is used.  This flame had one of
the the largest errors.  The same density, 50/50 Carbon/Oxygen flame
has $\MarksteinLength = -(1.028 \pm .003) \times 10^{-2} \ \mathrm{cm}$
when all the data is used, and $\MarksteinLength = -(1.047 \pm .002)
\times 10^{-2} \ \mathrm{cm}$ when the outward propagating only data
is used.  Figure \ref{fig:astro_outward} shows the fits for these two
flame when the outward data only is used.  They can be compared to
Figures \ref{fig:cf88flamecarbon} and \ref{fig:cf88flamecarbonoxygen}.
In the chemical combustion literature cited here, measurements are often
only made of outward-propagating flames (eg., \citealt{hassan98,kwon01}).

\begin{table}
\small
\begin{center}

\begin{tabular}{rcrrrrdd}
\tableline
\tableline
$\rho$ ($\gcc$)& $X_C$ & $S_l^0$ (cm s$^{-1}$) & $l_f^{0(I)}$ (cm) & $l_f^{0(II)}$ (cm) & \multicolumn{1}{c}{$\MarksteinLength$ (cm)} & \multicolumn{1}{c}{$\Markstein ^{(I)}$} & \multicolumn{1}{c}{$\Markstein ^{(II)}$} \\
\tableline
$2.5\times 10^7$ & 1.0 & $1.71 \times 10^5 $ & $6.53\times 10^{-2}$ & $3.86\times 10^{-2}$ & $-(3.107\pm.010)\times10^{-2}$ & -0.476 & -0.805 \\
\tableline
$5\times 10^7$  & 0.5 & $9.11 \times 10^4 $ & $5.28\times 10^{-2}$ & $3.07\times 10^{-2}$ & $-(5.496\pm.007)\times10^{-2}$ & -1.04 & -1.79 \\
$5\times 10^7$  & 1.0 & $6.81 \times 10^5 $ & $9.01\times 10^{-3}$ & $5.49\times 10^{-3}$ & $-(4.837\pm.038)\times10^{-3}$ & -0.537 & -0.881 \\
\tableline
$1 \times 10^8 $ & 0.5 & $2.92 \times 10^5$ & $9.93 \times 10^{-3}$ & $6.84 \times 10^{-3}$ & $-(1.028\pm.003)\times10^{-2}$ & -1.03 & -1.50 \\
$1 \times 10^8 $ & 1.0 & $2.13 \times 10^6$ & $1.85 \times 10^{-3}$ & $1.31 \times 10^{-3}$ & $-(8.366\pm.238)\times10^{-4}$ & -0.452 & -0.639 \\
\tableline
$2 \times 10^8 $ & 0.5 & $8.19 \times 10^5$ & $2.33 \times 10^{-3}$ & $1.72 \times 10^{-3}$ & $-(2.180\pm.014)\times10^{-3}$ & -0.935 & -1.27 \\
$2 \times 10^8 $ & 1.0 & $5.56 \times 10^6$ & $4.63 \times 10^{-4}$ & $3.74 \times 10^{-4}$ & $-(9.073\pm2.44)\times10^{-5}$ & -0.196 & -0.243 \\
\tableline
\end{tabular}

\end{center}
\caption{CF88 flame data.}
\label{table:cf88flamedata}
\end{table}

\begin{figure}
\plottwo{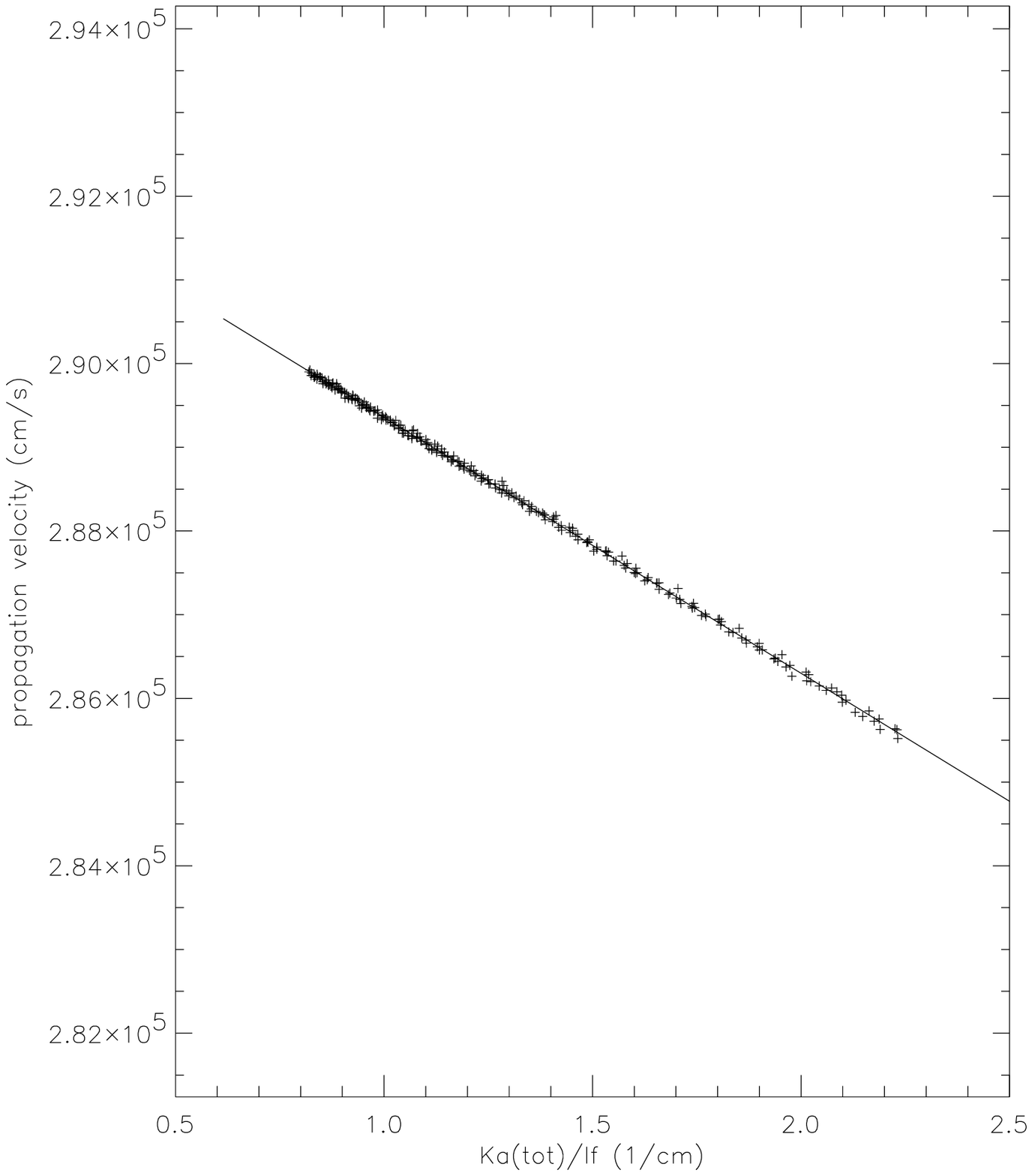}{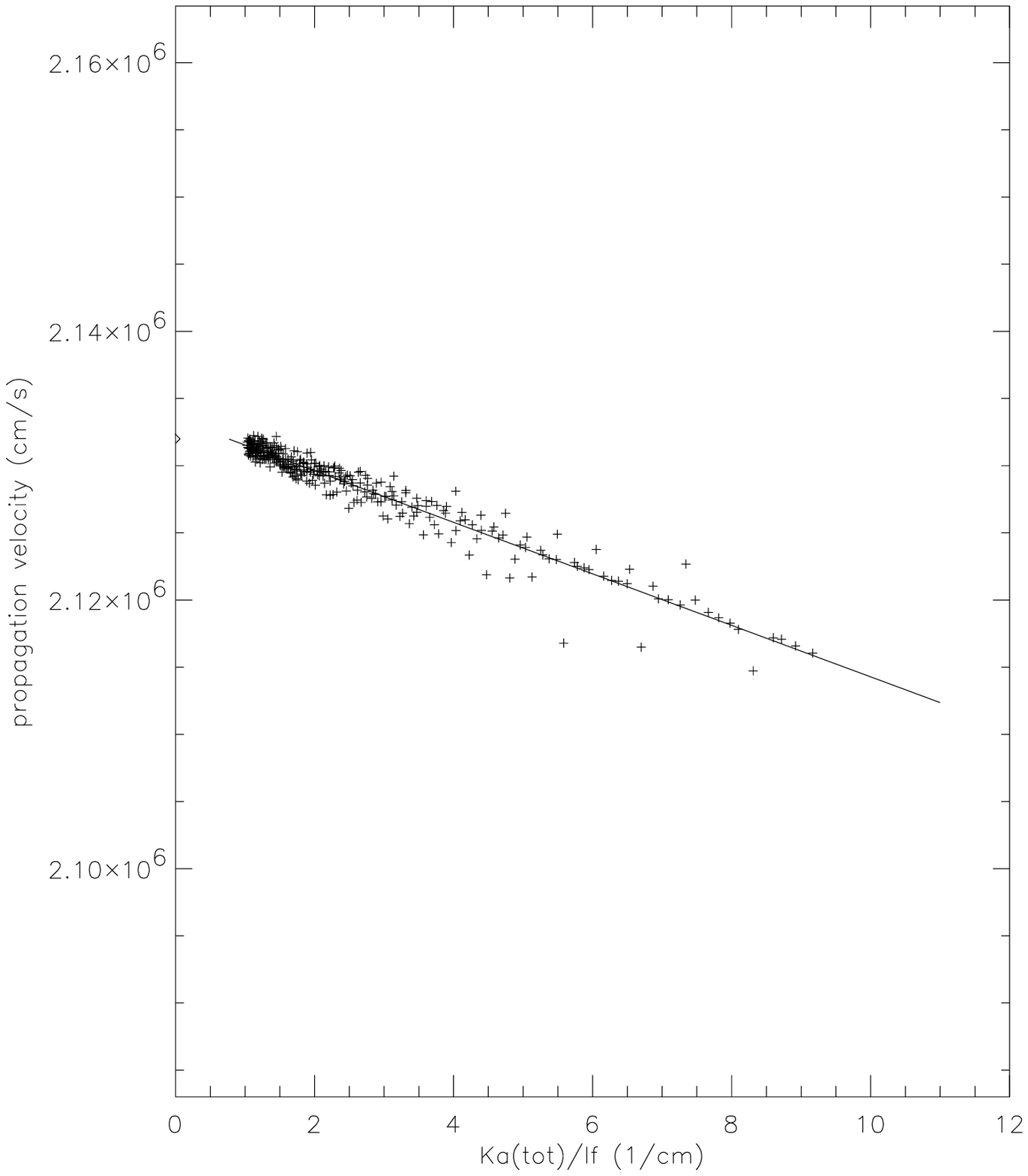}
\caption{\label{fig:astro_outward} 
         Fits of the $1\times 10^8 \gcc$ 0.5/0.5 Carbon/Oxygen (left)
         and pure carbon (right) flames using only the outward
         propagating flame data.}

\end{figure}

To assess the effects of the boundary conditions on the results,
we ran some flames with a domain twice the size as that used
in the above results.  Fig.~\ref{fig:cf88_long} shows the speed
vs. $\Karlovitz /l_f$ $1\times 10^8 \gcc$ pure carbon flame, where the
domain size was 7.68 cm, compared to 3.84 cm used in the main study
(Fig.~\ref{fig:cf88flamecarbon}).  The Markstein length computed for
this run is $\MarksteinLength = -(8.159 \pm .248) \times 10^{-4} \mathrm{cm}$, compared with the smaller domain value of $\MarksteinLength
= -(8.366 \pm .238) \times 10^{-4} \ \mathrm{cm}$.  We see that these
results are consistent within their fit errors.

\begin{figure}
\plotone{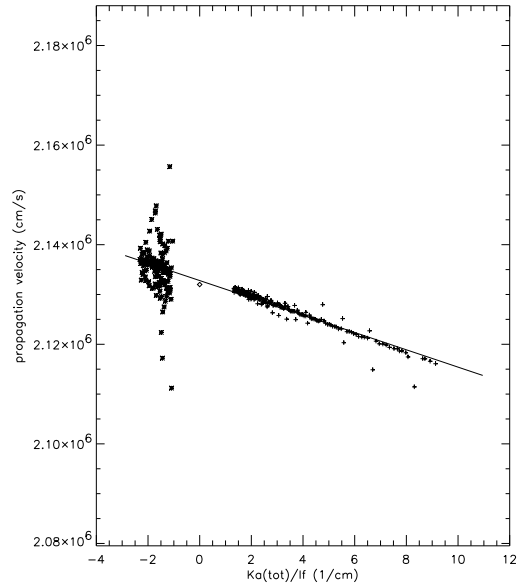}
\caption{\label{fig:cf88_long} 
         Results of the $1\times 10^8 \gcc$ pure carbon flame run with
         a domain twice the size used in the main study.  This can be compared
         with Fig. \ref{fig:cf88flamecarbon}.}
\end{figure}

The tables show that the Markstein length increases as the density or
carbon fraction decrease.  The reason for this is the predicted
dependence on Zeldovich number.  Table~\ref{table:cf88dataother} gives
the Zeldovich numbers for each of the CF88 flames, found by fitting an
Arrhenius rate to the measured $\enuc$ from the simulation.  A very
good Arrhenius fit can be found for each flame, but the parameters
vary from flame to flame.  $T_a$, for instance, varies slightly but
only by about 10\%.  If we assume that $T_a$ is roughly constant and
remember $T_b \gg T_u$ for these flames (see for instance
Fig.~\ref{fig:cf88flamecarbon}), then $\Zeldovich \approx T_a/T_b$, so
that increasing $T_b$ decreases $\Zeldovich$.  $T_b$ can be increased
by increasing the fuel (Carbon) fraction so more energy is released or
by increasing the density of the fuel and therefore, its degeneracy.
(The more degenerate the fuel, the less of the energy released by
burning will go into $P dV$ work, and so the temperature of the ashes
must increase instead).  We see that the measured Zeldovich numbers do
indeed follow these trends.

The relationship between $\Zeldovich$ and $\Markstein$ is further
explored in Fig.~\ref{fig:MaZeCF88}.   We see that there is a trend of
increasingly negative Markstein number with Zeldovich number, but there
is at least one other variable involved as well; different carbon
fractions give different trends.   Also, as degeneracy lifts 
(for the $\rho = 2.5 \times 10^{7} \gcc$ pure carbon flame), the
relationship grows more complicated.

\begin{figure}
\plotone{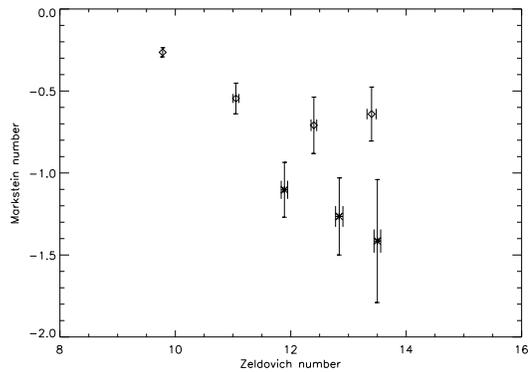}
\caption{\label{fig:MaZeCF88} 
         Plot of the calculated Zeldovich number for the six astrophysical
         flames considered here and their measured Markstein numbers.
         The vertical error bars reflect the values $\Markstein^{(I)}$ and
         $\Markstein^{(II)}$, with the symbols marking the average of
         the two for a given flame.  The bars with a star symbol (`$\ast$')
         represent the data from the 50\% carbon flames, and 
         those with a diamond symbol (`$\diamond$') represent the 100\% 
         carbon flames.  Horizontal error bars represent the uncertainty
         in measuring the Zeldovich number from fitting the burning rate.
    }
\end{figure}

\section{CONCLUSIONS}
\label{sec:conclusions}
We have reported on the curvature behavior of astrophysical and model
flames.   The behavior of astrophysical flames can be understood in
terms of simple geometry; a flame bent outwards towards fuel will burn
more slowly, because the diffusive heat transport is `diluted', 
whereas a flame bent inwards towards ash burns more quickly, because
heat transport is concentrated.   Quantitatively, we have seen that the
astrophysical flames reported on have a reaction to a range of
curvature- and flow-induced strain that is linearly related to the
magnitude of the strain for the small strains ($\Karlovitz_T < 1$)
examined here and have a Markstein length of size comparable to the
flame thickness.

Simplified model flames can have very different behaviors than the
astrophysical flames, if their structure is very different.  In
particular, a peaked burning rate and a separation between a preheat
zone and a burning zone are responsible for astrophysical flame
behavior on the scales explored here.   To measure small-scale flame
responses, then, real fully resolved flames must be calculated, or
models which take into account strain effects \citep{matalonmatkowsky}
must be used.   Methods which use advection-diffusion equations but do
not fully resolve the flame \citep{khoklov95} may work well at large
scales, but would be inappropriate to use for modelling small-scale
flame behavior.

We have demonstrated that to measure accurate burning velocities one
must consider the bulk burning rate, rather than using differencing on
the flame's position.  It is difficult to accurately determine
and subtract the fuel velocity, and relatively easy to accurately
compute integral quantities.  Even measuring the flame position
required care, as small uncertainties in the measured position can
result in significantly different computed Markstein numbers.

The magnitude of the response to strain increases with increasing
Zeldovich number.   For the flames examined here, decreasing carbon
fraction or decreasing density both increase the Zeldovich number.
However, the absolute magnitude of the response seen here has a
stronger dependence on composition of the fuel than through
the Zeldovich number.   The magnitude of the response is also smaller
than would be implied by the basic theory described in
\S\ref{sec:theory}.  More complete derivations, including the decrease
in density across the flame, have been described \citep{sun99}, but the
density jumps in the flames reported here are quite modest.   Other
fluid properties (temperature, conductivity, diffusivity) change a
great deal across the flame, more than the density jump would imply,
because of the degenerate EOS; however, most theory in the terrestrial
combustion literature assumes an ideal gas EOS.   These results, then,
must be extended to the degenerate EOS applicable to white dwarf
interiors.

The Markstein lengths measured here represent the smallest scale
wrinkling one can expect to see in astrophysical thermonuclear
flames.   Thus, it marks the lower scale end of the fractal-like
behavior one might expect to see in large-scale Landau-Darrieus
instability growth, as described in \citep{blinnsaswoos,ldfractal}.

A velocity law such as in Eq.~\ref{eq:marksteindef} implies a
burning velocity of zero --- that is, quenching --- for $\Karlovitz =
-1/\Markstein$ or $K = -S_l^0 / \MarksteinLength$.   Clearly,
extrapolating the behavior of the flames from the strain rates
investigated in this paper to strains sufficient to quench the flame is
problematic, but keeping this in mind, we can use this as an estimate
for the sorts of strain rates which can significantly affect local
burning rates.  A periodic sinusoidal shear flow of amplitude $A$ and
wavelength of $n$ flame thicknesses has a peak strain rate of $(2 \pi
A)/(n l_f)$.    For that to be equal to the estimated quenching strain
above, $(A/S_l^0) = n/(2 \pi \Markstein)$.    Since the flames
investigated had $\Markstein \approx 1$, we see that periodic strains
on the flame thickness scales with velocities comparable to the flame
speed can be expected to significantly alter the flame's burning rate.
See for example, \citet{flamevortex}.

The direction of the astrophysical flame response to strain is to act
against the strain, thus partially stabilizing an initially planar
flame against wrinkling by strain or instabilities.   Given the
measurements presented here, we can predict the flame's behavior under
the Landau-Darrieus instability.   From Eq.~\ref{eq:marksteinld} we can
calculate the change in the growth rate of the instability, and the
largest mode stabilized, by the non-constant flame velocity.  If, as in
\S\ref{sec:theory}, $k_c = (\alpha-1)/(2\alpha l_f
|\Markstein|)$ is the critical wavenumber for stability, we can
calculate the largest wavelength stabilized, $\lambda_c = 2 \pi /
k_c$.  These are listed, for the astrophysical flames studied here, in
Table~\ref{table:cf88ld}.   Results are given in terms of flame
thicknesses (we use method I for concreteness) and in centimeters.  We
note that the increase in Markstein number as one goes to lower density
approximately cancels the expected decrease in stability due to
increasing density contrast as degeneracy lessens.

\begin{table}
\begin{center}

\begin{tabular}{ccrr@{$\, \pm \,$}l@{}l}
\tableline
\tableline
$\rho$ ($\gcc$)& $X_C$ & $\lambda_c/l_f^{0(I)}$ & \multicolumn{3}{c}{$\lambda_c$ (cm)}\\
\tableline
$2.5 \times 10^7$  & 1.0 & 11.1 & (7.24 & .02) &$\times 10^{-1}$ \\
\tableline
$5 \times 10^7$  & 0.5 & 50.8 & (2.15 & .003) &$\times 10^{0}$ \\
$5 \times 10^7$  & 1.0 & 14.3 & (1.28 & .01 ) &$\times 10^{-1}$ \\
\tableline
$1 \times 10^8 $ & 0.5 & 48.9 & (4.86 & .01 ) &$\times 10^{-1}$ \\
$1 \times 10^8 $ & 1.0 & 13.9 & (2.57 & .07 ) &$\times 10^{-2}$ \\
\tableline
$2 \times 10^8 $ & 0.5 & 52.3 & (1.21 & .007) &$\times 10^{-1}$ \\
$2 \times 10^8 $ & 1.0 &  8.46& (3.91 & 1.07) &$\times 10^{-3}$ \\
\tableline
\end{tabular}

\end{center}
\caption{Predicted critical wavelength for Landau-Darrieus instability,
in terms of flame thickness (method I) and in centimeters, for the CF88
flames described here.}
\label{table:cf88ld}
\end{table}

Along with the growth rate of the Landau-Darrieus instability, the
curvature behavior of the flame also affects its non-linear stabilization.
As is well known \citep{zeldovich85}, the growth of a perturbation in a
propagating interface may be limited by cusp formation.  This is not a
property of the Landau-Darrieus instability, but simply of a propagating
interface by Huygens's principle (see, for example, \citealt{hechtoptics})
and can be demonstrated with the so-called `G-equation' (see for instance
\citealt{lawsung}).   There has been some astrophysical interest in
this very recently \citep{MPA03}.   Since cusp formation involves large
curvatures, is very strongly affected by the curvature behavior of the
flame, and indeed, cusp formation can be completely inhibited by the
flame behavior, as shown in Fig.~\ref{fig:geqn-cusping}.

\begin{figure}
\plotone{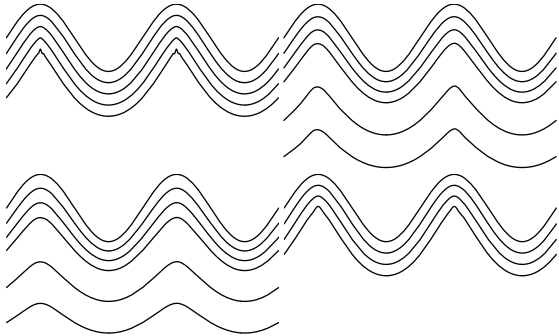}
\caption{\label{fig:geqn-cusping} 
         Solutions to the `G equation', $\partial G/\partial t =
         |\nabla G|(1 + \MarksteinLength \partial^2 G/\partial x^2)$
         tracking a propagating surface through a domain with no flow.
         The interface is originally a sinusoid of dimensionless
         amplitude 1, and wavelength $2 \pi$, and propagates
         downwards.  If the flame propagates at a constant velocity of
         1, it cusps as shown in the top left frame (at times $0, 1/3,
         2/3, 1, 4/3$).  If the interface has a negative Markstein
         length equal to $1/20$, the results are shown in the top
         right (at times $0, 1/3, 2/3, 1, 2, 3$) and we see that
         cusping is postponed.  Doubling the Markstein length gives
         behavior as shown on the bottom left, at times $0, 1/3, 2/3,
         1, 2, 3$.  Cusping is suppressed, and the flame begins to
         flatten.  On the other hand, a positive Markstein length of
         $1/20$ enhances cusping as shown on the bottom right (shown
         at times $0, 1/3, 2/3, 1$.)  Numerical solutions of the PDE
         were calculated using Mathematica.  }
\end{figure}

The local flame response to shear measured here makes it harder to
significantly accelerate the burning on scales of a few tens of flame
thicknesses by instabilities, and may not strongly affect larger-scale
wrinkling.  The same local flame response may, however, make a
transition to detonation easier, as the burning rate changes
significantly with shears of order $S_l^0/l_f^0$.

This work has focused on single-step reaction mechanisms.   Realistic
flames, both astrophysical and terrestrial, involve many reactions
and intermediate species.  Future work will extend the results presented
here to such multi-stage reaction networks.

\acknowledgements 

The authors gratefully acknowledge Alan Kerstein for many instructive
discussions on flames.  Support for this work was provided by DOE
grant number B341495 to the ASCI/Center for Astrophysical
Thermonuclear Flashes at the University of Chicago and the Scientific
Discovery through Advanced Computing (SciDAC) program of the DOE,
grant number DE-FC02-01ER41176 to the Supernova Science Center/UCSC.
LJD is supported by the Krell Institute CSGF.  All simulations were
performed with the \FLASH code, version 2.2.  Some calculations were
performed on the UCSC UpsAnd cluster supported by an NSF MRI grant
AST-0079757.  The astrophysical flame models presented in the appendix
are available for comparison at
\url{http://www.ucolick.org/~zingale/flame_models/}.  The \FLASH code
is freely available at \url{http://flash.uchicago.edu/}.

\appendix
\section{VERIFICATION TESTS}
\label{sec:appendix}

In this paper, we describe results from the \FLASH code with additional
modules for modelling diffusion and conductivity, and for computing
hydrodynamics in 1-d spherical coordinates.   The \FLASH code has been
well tested elsewhere without these modules \citep{flash,vandv}; in
this appendix we present test results of the new modules added for the
simulation of flames.

\subsection{1-D Spherical Hydro}

Since most of the simulations presented here were performed in
spherical coordinates, we first test our solutions of the equations of
hydrodynamics in spherical coordinates.  The hydrodynamic algorithm
and implementation used in the \FLASH code have been well-tested in
Cartesian coordinates, so we need only test that the area factors and
adaptive mesh refinement (AMR) routines are correctly implemented for
spherical coordinates in the \FLASH code.  Shown in
Fig.~\ref{fig:sedov_sph} is the result of a Sedov explosion
\citep{sedov} propagating in a 1-d spherically-symmetric domain.  This
calculation included AMR.  The calculated result is shown with
circles, and the predicted density profile \citep{landaulifshitz} is
shown with a solid line.  Also plotted is the shock position versus
time.  We see that the shock is well modeled with the \FLASH code in
spherical coordinates.

\begin{figure}
\plottwo{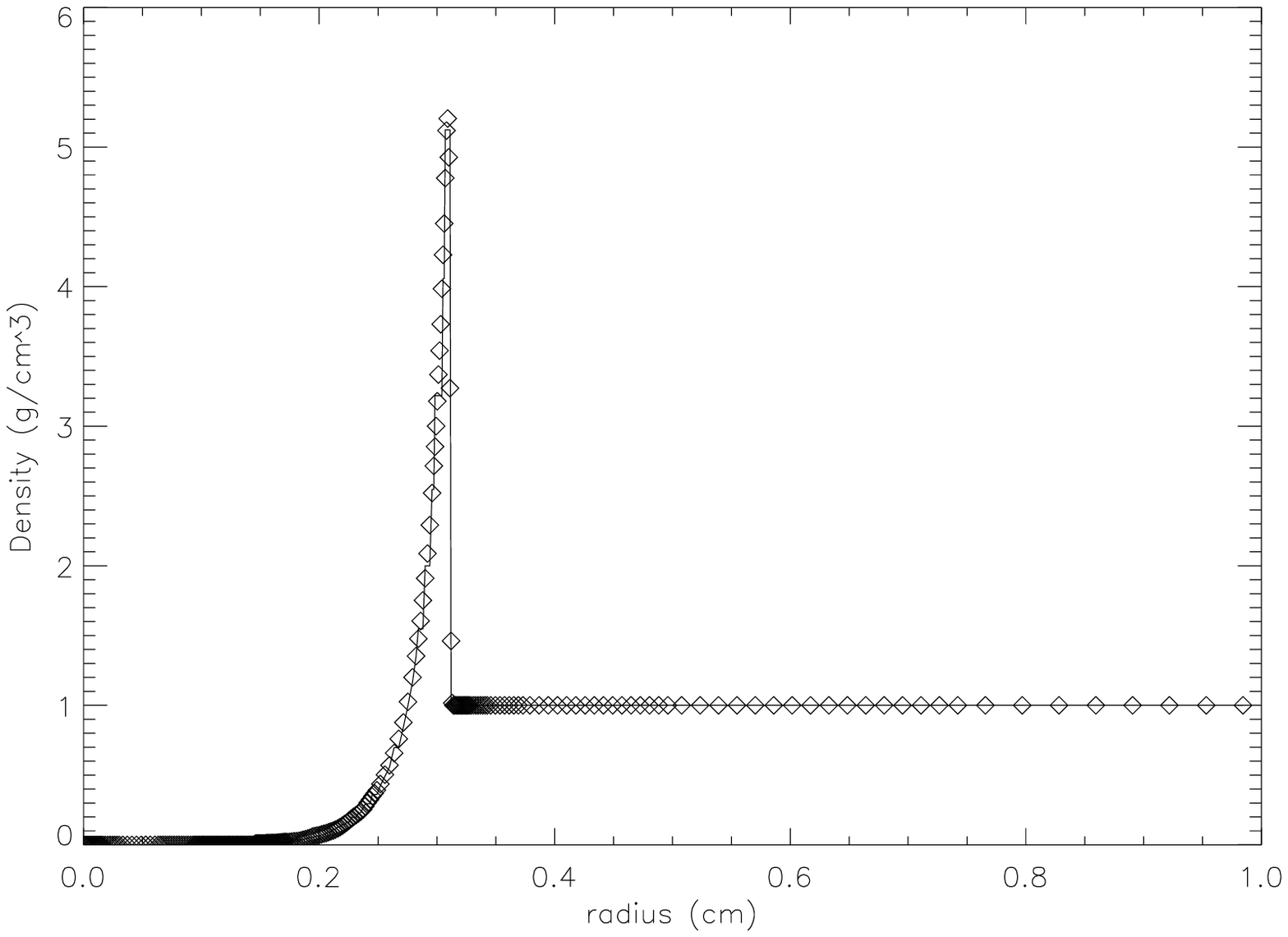}{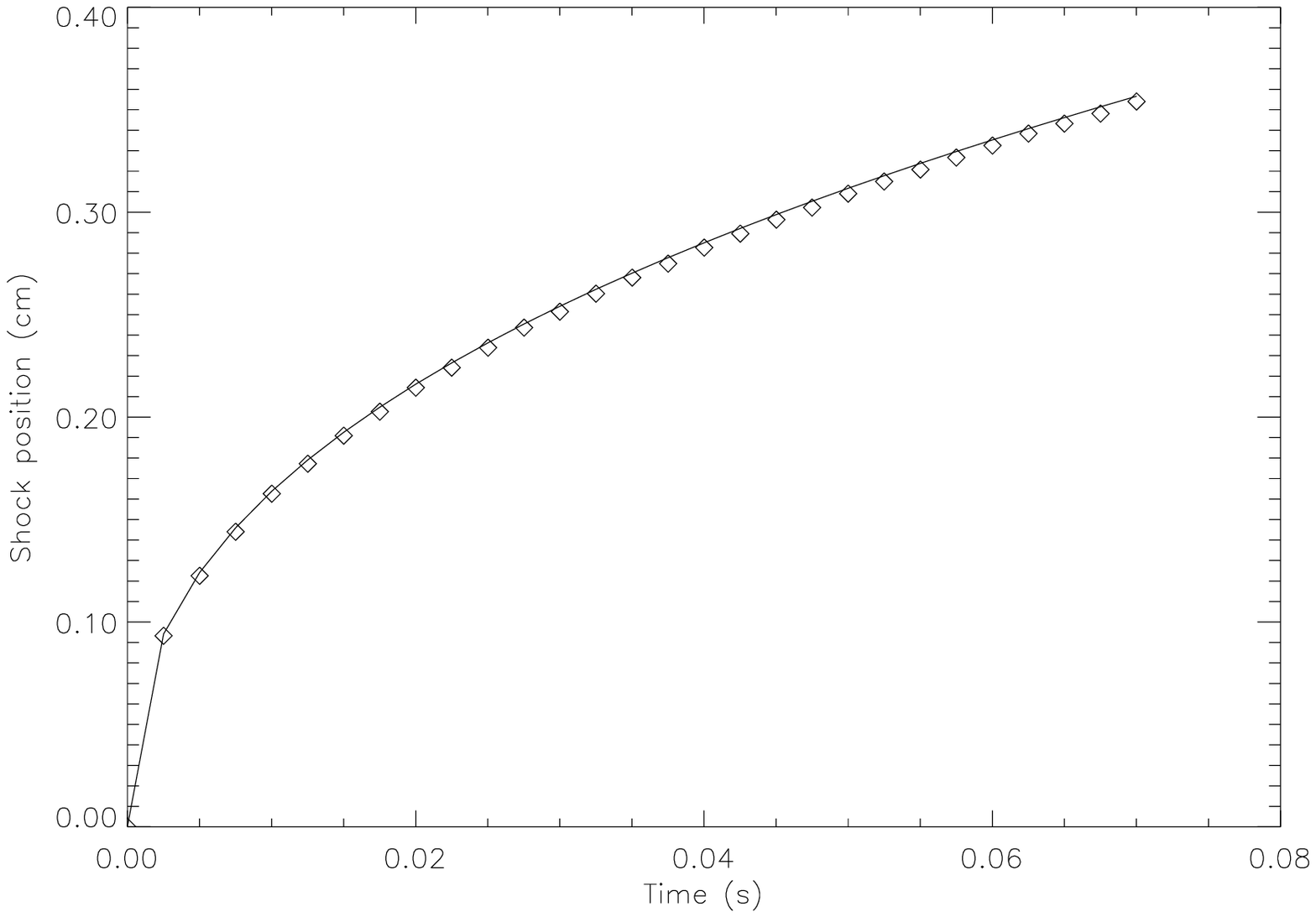}
\caption{\label{fig:sedov_sph} 
	  On the left is the density profile of a spherically expanding
	  Sedov explosion 0.05 seconds after the explosion at $r=0$.
	  The explosion energy was 1 erg, into an ambient pressure of
	  $10^{-5} {\ \mathrm{dynes \ cm^{-2}}}$ and an ambient density of $1
	  \gcc$.   On the right is the shock position of
	  the same simulation over time.  Simulated results are shown
	  with open circles, and predicted results with a solid line.
}
\end{figure}

\subsection{Diffusion Tests}

To test the diffusion module described in \S \ref{sec:methods:diff},
we diffuse a Gaussian pulse of temperature through a
constant-density domain.  As is well known, a planar initially Gaussian
pulse will remain a Gaussian when diffused with a constant diffusivity 
$D_{\mathrm{th}}$, with a width
\begin{equation}
\sigma(t) = \sqrt{\sigma_0^2 + 2 D_{\mathrm{th}} (t - t_0)},
\end{equation}
where $\sigma_0 = \sigma(t = t_0)$.
We use this to test the thermal diffusion in Cartesian coordinates.  An
initial temperature pulse is created and evolved in the \FLASH code.
For this test, hydrodynamics is not calculated, so that we can get a
clean test of the diffusion.

Quantitative results are shown in Fig.~\ref{fig:diffuse-tempwidths}.
The measured Gaussian widths were computed by fitting the evolved
profile with a Gaussian of arbitrary amplitude and width.  As we see,
we get excellent agreement over a diffusive timescale.   The same test
was run with a species diffusion problem; identical results were
obtained, since the code for the two diffusion operators is identical.

\begin{figure}
\plottwo{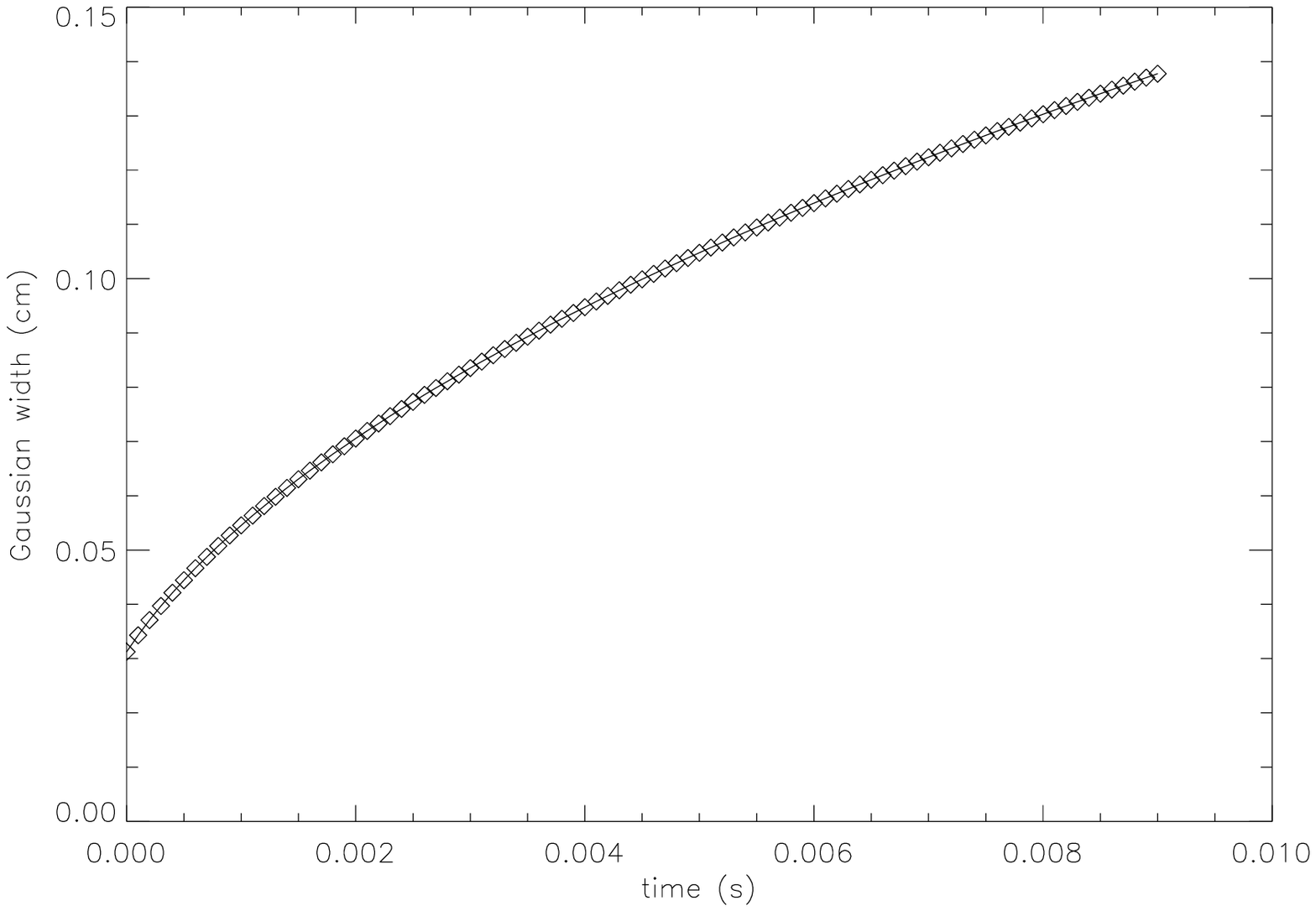}{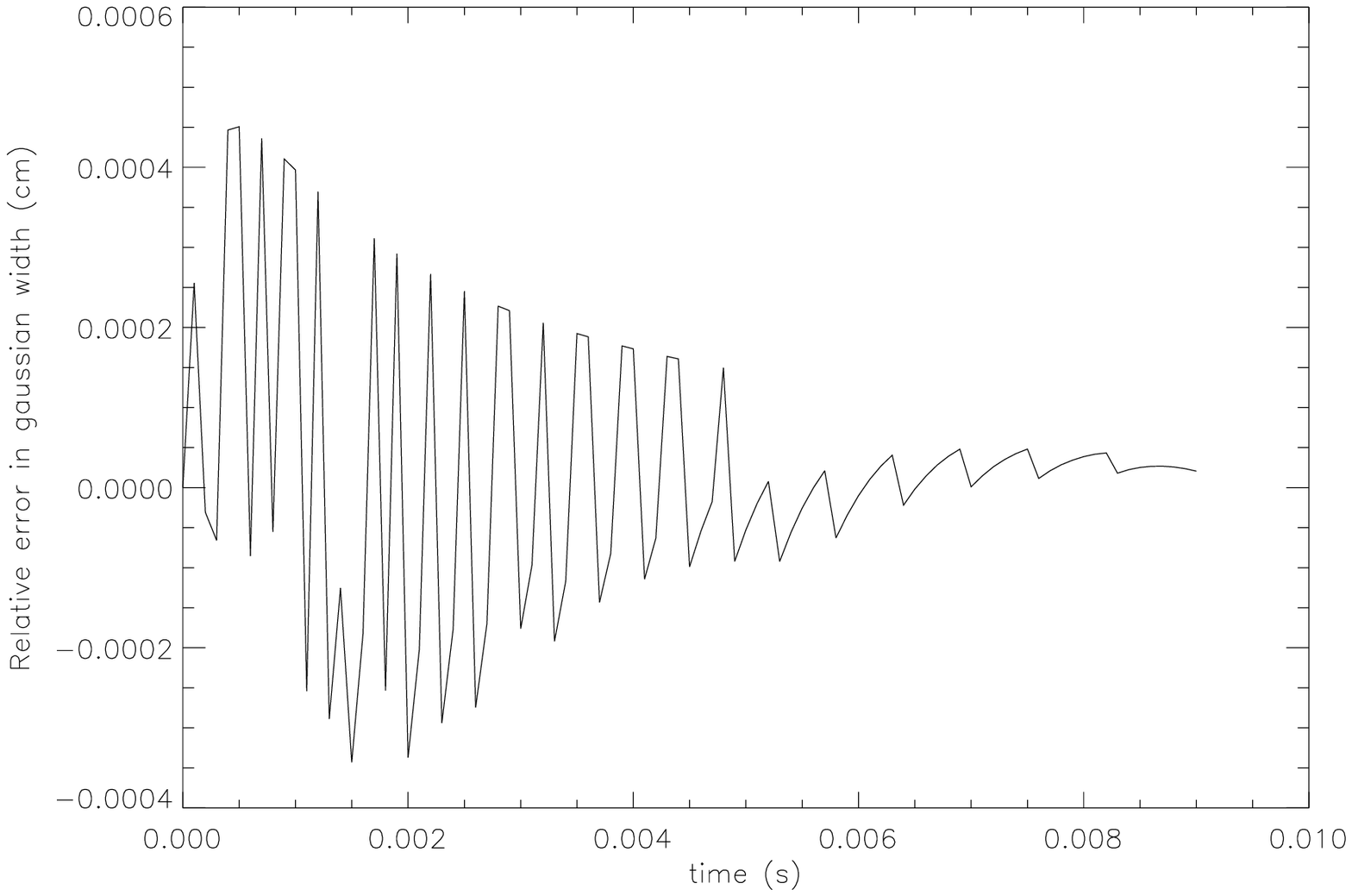}
\caption{\label{fig:diffuse-tempwidths} 
	  On the left, the evolution of the best-fit Gaussian widths as
	  a function of time in the simulation, both predicted (solid
	  line) and measured (`$\diamond$'), for 
	  a constant diffusivity of $D_{\mathrm{th}} = 1 {\ \mathrm{cm^2 \ s^{-1}}}$.  
          The original profile is a Gaussian of
	  width $\sigma = 1/32 {\ \mathrm{cm}}$; the resolution
	  is $\Delta x = \sigma_0 / 16 = 1/512 {\ \mathrm{cm}}$.
          Plotted on the right is
	  the relative difference between the best-fit Gaussian width
	  and the predicted Gaussian width.  
} 
\end{figure}

A 1-d spherically symmetric Gaussian profile is not a solution
of the spherically-symmetric diffusion equation
\begin{equation}
\label{eq:sphericaldiffusion}
\frac{\partial T}{\partial t} = -\frac{1}{r^2} \frac{\partial}{\partial r}
  r^2 \frac{\partial}{\partial r} D_{\mathrm{th}} T.
\end{equation}
Instead, the solutions are spherical Bessel functions.  Thus we
consider a temperature profile
\begin{equation}
\label{eq:sphericaldifftest}
T(r,t) = T_0 ( 1 + j_0(k r) e^{-k^2 D_{\mathrm{th}} t} ).
\end{equation}

We simulated a diffusing Bessel function profile with $k =
1~{\mathrm{cm^{-1}}}$, $T_0 = 1.~{\mathrm{K}}$, and $D_{th} =
100~{\mathrm{cm^2~s^{-1}}}$.   A close-up of the calculated and
predicted profiles and the error after 2 diffusion times ($0.02$
seconds) are shown in Fig.~\ref{fig:spherediffuse-late}.  The error is
small everywhere except at the right boundary, because the
infinite domain in which the spherical Bessel functions form a solution
to the diffusion equation is modeled by a truncated domain with a
zero-gradient boundary condition.  Reflecting boundary conditions are
used at the left boundary.

\begin{figure}
\plottwo{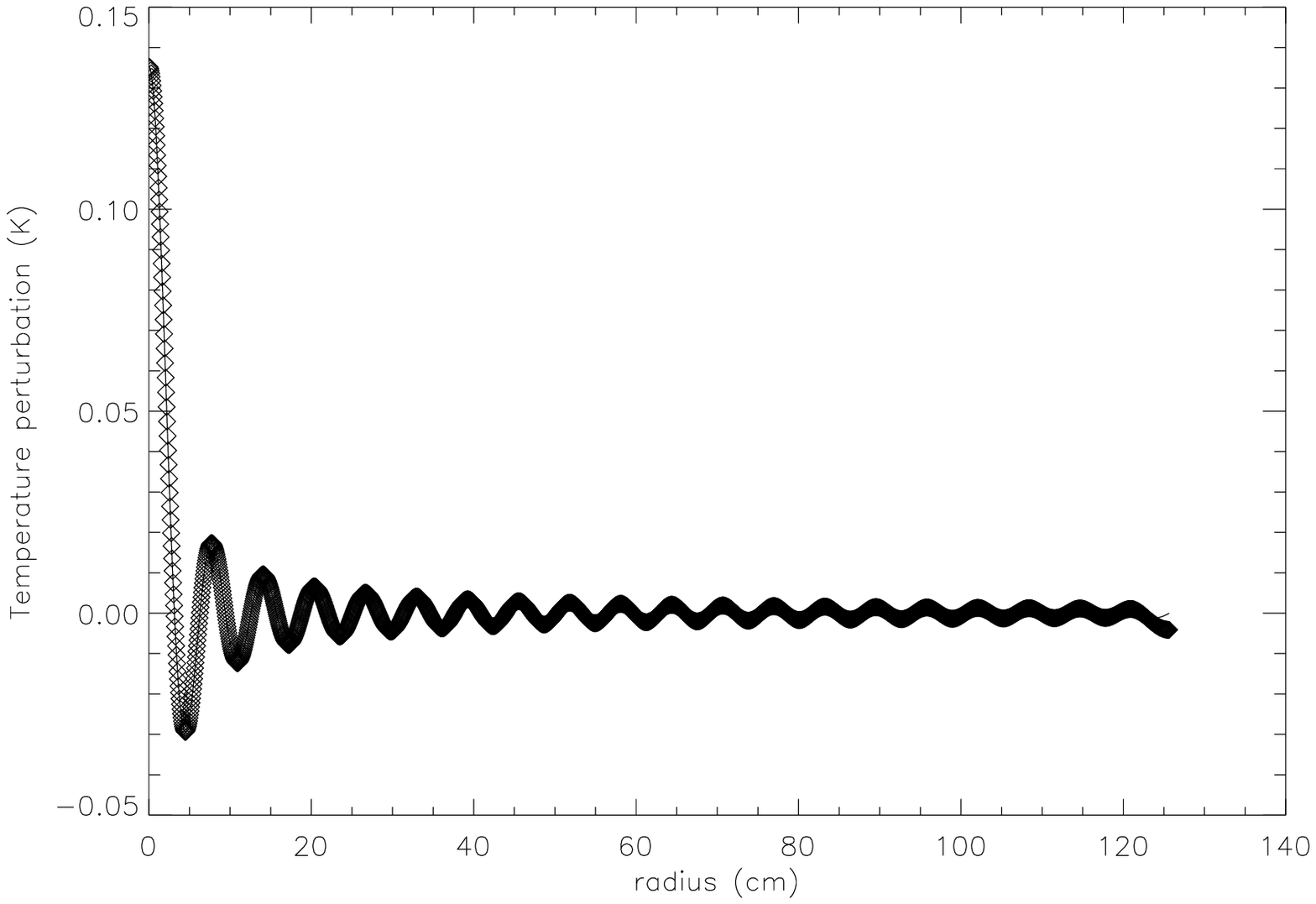}{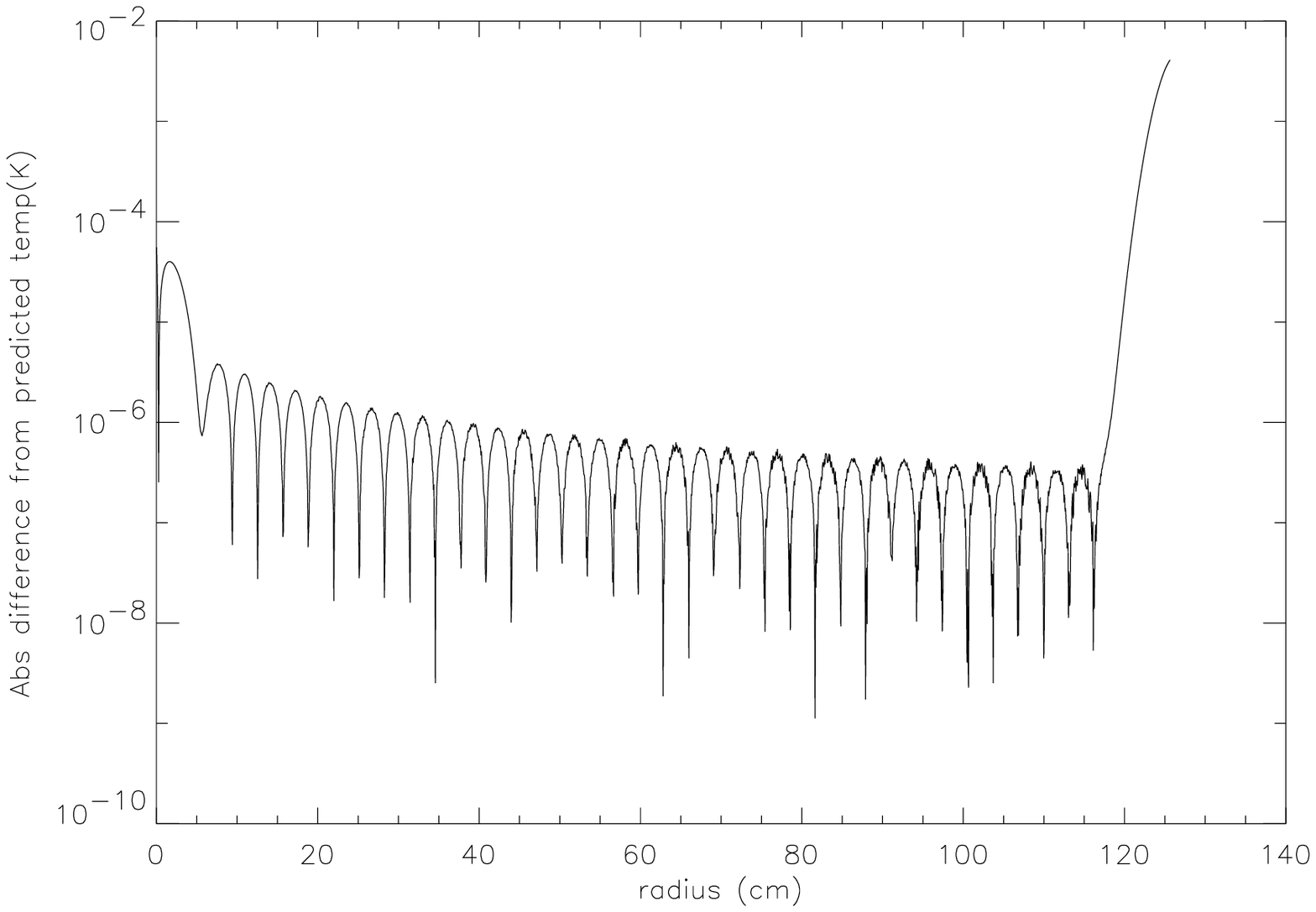}
\caption{\label{fig:spherediffuse-late} 
	  On the left, the temperature perturbation after two diffusion times
          (`$\diamond$'), and the predicted profile (solid line).  On the right,
          the error in the evolved temperature perturbation at the same 
          time.   Error is small everywhere except near the boundary 
          conditions.
}
\end{figure}

\subsection{Boundary Conditions}

Verifying a boundary condition is difficult; here we simply
present evidence that the boundary condition allows the material
from the flame to smoothly leave the domain.   Fig.~\ref{fig:lagrangebc}
shows a space-time diagram of points in the domain of a planar
flame simulation described in \S\ref{sec:experiments}.   The
Lagrangian trajectories were calculated as a post-processing step
on the results of the simulations, taking a set of initial points
and tracking them through the velocity field.   When a point
is overtaken by the flame, it expands and falls behind the flame,
eventually leaving the domain; material ahead of the flame moves
slowly in the other direction.  In both cases, the material must
advect smoothly out of the domain.   We see in Fig.~\ref{fig:lagrangebc}
that material advects out of the domain without any artifacts.
The initial conditions for this simulation was hot ($T = 0.96$) 
ash on the left up to $x=10$, cold ($T = 0.1, \rho = 1$) fuel on
the right, in pressure equilibrium, and $\Lewis = \infty$.  A
$\Zeldovich  \sim 5$ Arrhenius burning rate and a $\gamma = 5/3$
ideal gas EOS were used.  Ignition occurred after about 70 seconds.

\begin{figure}
\begin{center}
\plotone{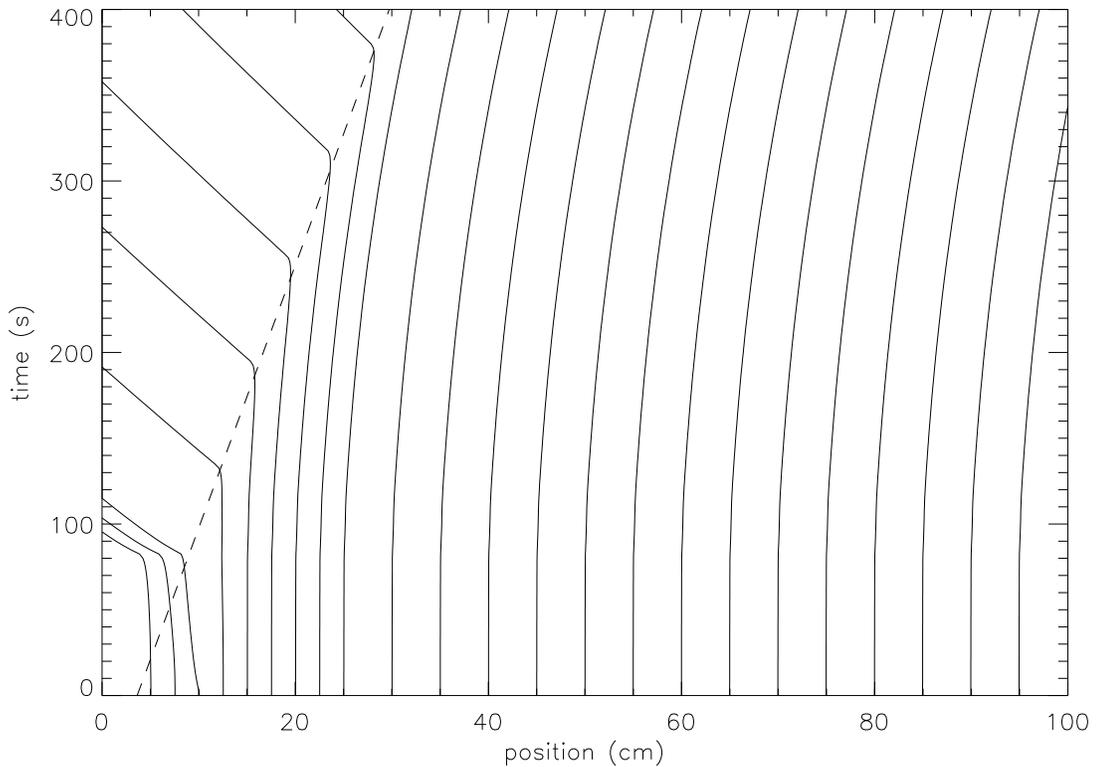}
\caption{\label{fig:lagrangebc}
	  Points moving along Lagrangian trajectories as a planar
          Arrhenius model flame
          ignites at $x = 10$ at time $t \approx 70$.  Material
          on both sides of the flame smoothly leaves the domain.}
\end{center}
\end{figure}

\subsection{Flame Speeds and Structure}
\label{sec:resstudy}
\subsubsection{Model Flames}

For our model flames, in Table \ref{table:flame_res_arr} we summarize
the results of a resolution study for the $\Lewis = \infty$,
$\Zeldovich \approx 5$ flame.  We see consistent resolution
requirements for both these flames and the astrophysical flames listed
above.   We use a finest resolution of $1.56 \times 10^{-2}
\ \mathrm{cm}$, which is seen to provide well-converged results, for the
$\Zeldovich \sim 5$ Arrhenius flames and the KPP flames described here,
even though the KPP flames and finite Lewis-number Arrhenius flames
are thicker than the infinite Lewis number Arrhenius flame.

\begin{table}
\begin{center}

\begin{tabular}{crc}

\tableline
\tableline
$\delta x$ & $n_{\mathrm{zones}}$ & $S_l^0$ \\
\tableline
$1.25 \times 10^{-1}$ & 7  & $4.046 \times 10^{-2}$ \\
$6.25 \times 10^{-2}$ & 14 & $4.001 \times 10^{-2}$ \\
$3.13 \times 10^{-2}$ & 28 & $3.999 \times 10^{-2}$ \\
$1.56 \times 10^{-2}$ & 57 & $3.999 \times 10^{-2}$ \\
\tableline

\end{tabular}

\end{center}
\caption{Numerically determined laminar planar flame
velocity as a function of resolution in the \FLASH code.  The
resolution is expressed as the zone size, and the number of zones
across the thermal thickness ($l_f^{0(I)}$) of the $\Lewis = \infty$,
$\Zeldovich \sim 5$ Arrhenius flame.}
\label{table:flame_res_arr} 
\end{table}

\begin{table}
\begin{center}

\begin{tabular}{crrrr}

\tableline
\tableline
$\delta x $ (cm) & $n_{\mathrm{zones}}$ & $S_l^0 \ \mathrm{cm s^{-1}}$ & $l_f^{0(I)}$ & $l_f^{0(II)}$ \\
\tableline
$5.21\times 10^{-5}$ & 34 & $2.135\times 10^5 $ & $1.75\times 10^{-2} $ & $1.22 \times 10^{-2}$ \\
$1.04\times 10^{-4}$ & 18 & $2.132\times 10^5 $ & $1.85\times 10^{-2} $ & $1.31 \times 10^{-2}$ \\
$2.08\times 10^{-4}$ & 10 & $2.162\times 10^5 $ & $2.04\times 10^{-2} $ & $1.51 \times 10^{-2}$ \\
$4.17\times 10^{-4}$ & 6  & $2.194\times 10^5 $ & $2.47\times 10^{-2} $ & $1.64 \times 10^{-2}$ \\
\tableline

\end{tabular}

\end{center}

\caption{
Numerically determined laminar planar flame velocity for the $1\times
10^8 \gcc$ pure carbon CF88 flame as a function of resolution.  The
resolution is expressed as the zone size and the number of zones
across the thermal thickness of the flame.  We see that the speed
converges at roughly 20 points in the laminar flame thickness.}
\label{table:flame_res} 
\end{table}

\subsubsection{CF88 flames}

The properties of astrophysical carbon flames were investigated in
great detail in \citet{timmeswoosley}.  There, several different
methods and nuclear reaction networks were tried.  All of the reaction
networks they considered involved multiple reactions.  The absence of
multiple reactions in our study makes the temperature structure
considerably simpler.  After the initial carbon burns, the temperature
remains flat.  With a more extensive network, later reactions would
continue to heat (or possibly cool) the region behind the flame.  The
10/90 flame width definition would be more sensitive to these later
reactions, and would tend to produce a wider flame than those we get.

Flames were simulated propagating through a range of densities
relevant to white dwarf interiors.  Of particular interest is
determining how much spatial resolution is required to accurately
simulate a flame.  The results presented here were used as the
benchmark for the flame propagation experiments presented in this paper.

Fig.~\ref{fig:cf88flameprofile} shows the structure of a $5\times
10^7 \gcc$ pure carbon flame, propagating to the left.  We see that the
large jump in temperature happens right at the peak nuclear energy
generation rate, and then the temperature is almost perfectly flat
behind that.  The temperature jump is accompanied by almost a factor
of two density drop behind the flame.

\begin{figure}
\plotone{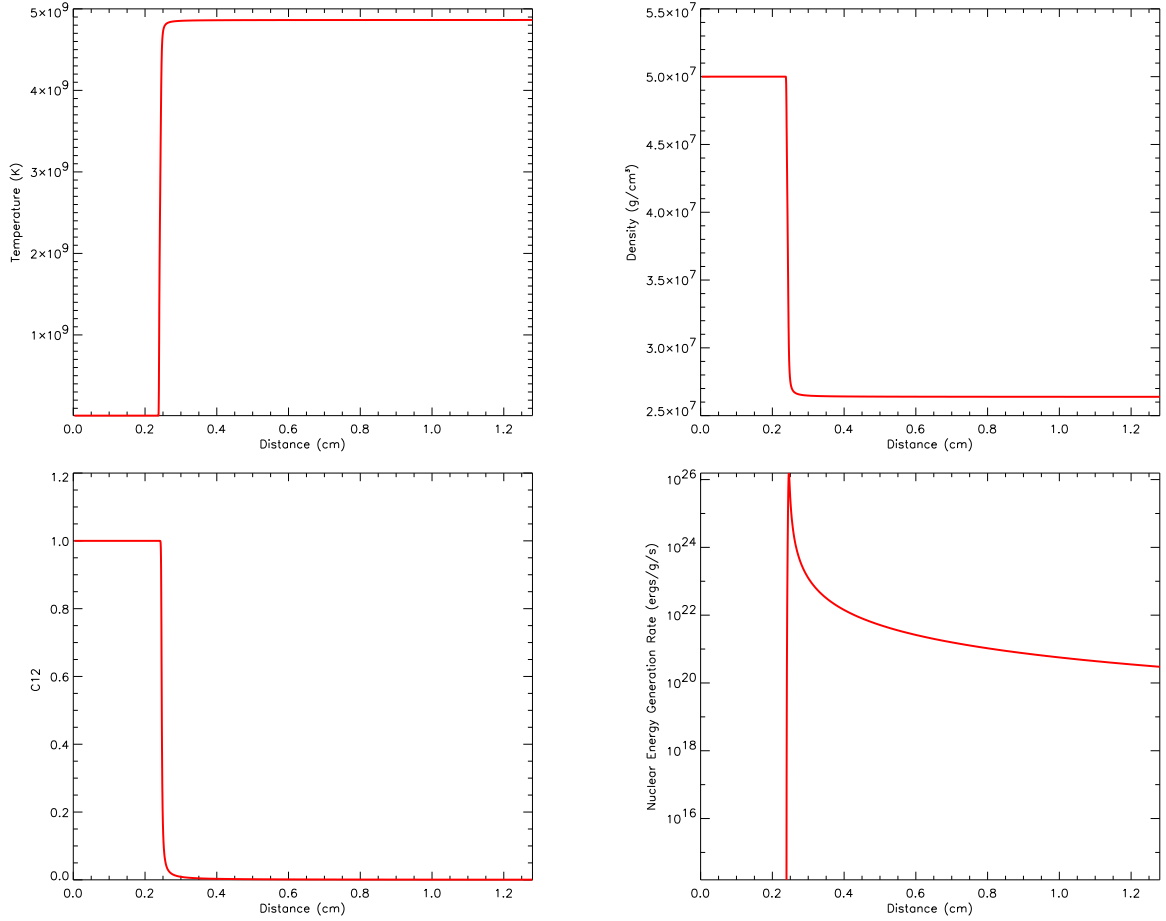}
\caption{\label{fig:cf88flameprofile} Flame structure for a planar $5\times 10^7 \gcc$ pure carbon CF88 flame.  Shown are temperature (top left), density (top right), carbon mass fraction (bottom left), and nuclear energy generation rate (bottom right).}
\end{figure}

An accurate description of the deflagration speed requires resolving
the front.  For all the flames studied here, we ran with several
different resolutions, in order to find out how much resolution was
necessary for the flame speed and width to converge.  Table
\ref{table:flame_res} summarizes the results of one such resolution study, 
for the $1\times 10^8 \gcc$ pure carbon flame.  We see that 20-40
zones in the flame width is sufficient to determine the flame speed
accurately.  The main calculations in this paper were run with a
resolution that put about 20 computational zones in the flame width
($l_f^{0(I)}$).  We did not consider simulations with less than 5
zones in the flame width.

\section{CONSERVATIVE INTERPOLATION}
\label{sec:appendixb}

In FLASH, the computational zones are organized into blocks, typically
containing 8-16 uniformly gridded zones in each coordinate direction.
Adapting to the flow is achieved by allowing blocks to vary in
resolution by a factor of two with respect to neighboring blocks.
When a region needs to be refined, the parent block is cut in half in
each dimension, creating $2^d$ new blocks, where $d$ is the
dimensionality.  Initializing the newly created zones in each child
block requires that we reconstruct the zone-average data in the parent
(and its neighbors) and use this to compute the child data in a
conservative fashion.  This method is also needed at jumps in
resolution, when exchanging guardcell information between neighboring
blocks.

Reconstruction is achieved by fitting a quadratic polynomial to the
parent data.  Since \FLASH a finite-volume code, the coefficients of
the interpolation polynomial must be chosen to reproduce the zone average values of
the parents in the stencil.  Figure
\ref{fig:sph_quad} shows the three coarse zones used to initialize the
children.

\begin{figure}
\plotone{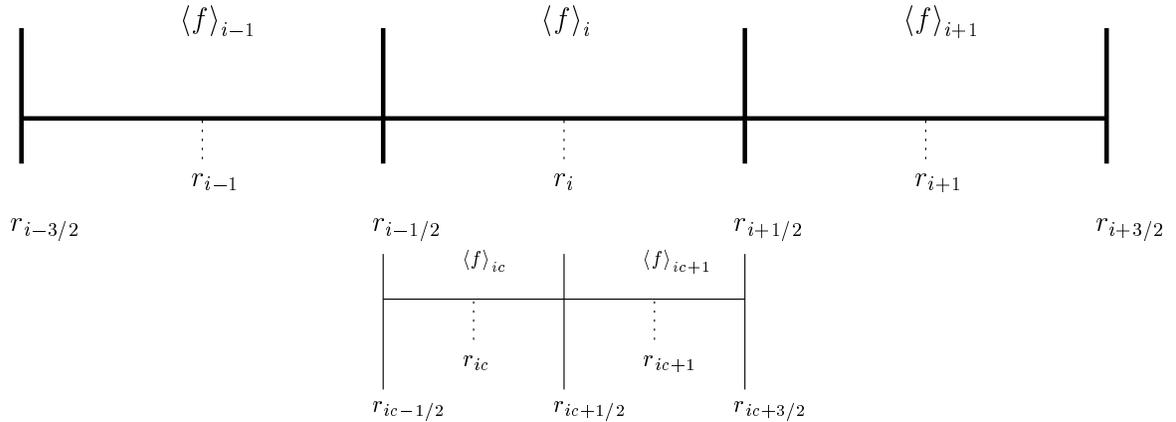}

\caption{\label{fig:sph_quad} Schematic showing three coarse zones with 
zone-averaged data $\avgsub{f}{i-1}$, $\avgsub{f}{i}$, and
$\avgsub{f}{i+1}$ (top).  The $i$-th zone is refined to create two
children (shown on bottom), whose data is to be initialized with
$\avgsub{f}{ic}$ and $\avgsub{f}{ic+1}$, through conservative
interpolation.}

\end{figure}

Given the polynomial,
\begin{equation}
\label{eq:reconstpoly}
f(r) = a_1 r^2 + a_2 r + a_3 \enskip ,
\end{equation}
we apply the constraints,
\begin{equation}
\label{eq:constraint}
\frac{4 \pi}{V_{i}} \int_{r_{i-1/2}}^{r_{i+1/2}} r^2 dr f(r) = \avgsub{f}{i} \enskip ,
\end{equation}
to each of the parent zones and solve to find $a_1$, $a_2$, and $a_3$.
Here, $V_i$ is the volume of the zone, which, for spherical coordinates, is
\begin{equation}
V_i = \frac{4}{3} \pi (r_{i+1/2}^3 - r_{i-1/2}^3) \enskip ,
\end{equation}
and $\avgsub{f}{i}$ is the average value of variable $f$ in zone $i$.
The zone edges are denoted by half-integer indices, as illustrated in
Figure \ref{fig:sph_quad}.
This polynomial can then be integrated over the children to initialize their
data.  It is also used when filling guardcells at coarse-fine interfaces.

\newcommand{\thetanote}[2]{\raisebox{-0.5ex}{\LARGE $\Theta$}_{#1}^{#2}\,}

Applying Eq. \ref{eq:constraint} to the $i$-th zone yields 
\begin{equation}
\label{eq:ithzone}
\frac{3}{5} a_1 \thetanote{3}{5} r_i + \frac{3}{4} a_2 \thetanote{3}{4} r_i + a_3 = \avgsub{f}{i}
\end{equation} 
where, for notational convenience, we define the operator
\begin{equation}
\thetanote{n}{m} r_i \equiv \frac{r_{i+1/2}^m - r_{i-1/2}^m}{r_{i+1/2}^n - r_{i-1/2}^n}
\end{equation}
We also use the $i-1$ and $i+1$ zones, and get
\begin{equation}
\label{eq:im1zone}
\frac{3}{5} a_1 \thetanote{3}{5} r_{i-1} + \frac{3}{4} a_2 \thetanote{3}{4} r_{i-1} + a_3 = \avgsub{f}{i-1}
\end{equation}
\begin{equation}
\label{eq:ip1zone}
\frac{3}{5} a_1 \thetanote{3}{5} r_{i+1} + \frac{3}{4} a_2 \thetanote{3}{4} r_{i+1} + a_3 = \avgsub{f}{i+1}
\end{equation}

Equations \ref{eq:ithzone}--\ref{eq:ip1zone} can be solved for $a_1$, $a_2$, and $a_3$ to yield:
\begin{equation}
a_2 = \frac{4}{3} 
      \frac{\left [ \thetanote{3}{5} r_i - \thetanote{3}{5} r_{i+1} \right ]
            \left ( \avgsub{f}{i} - \avgsub{f}{i-1} \right ) -
            \left [ \thetanote{3}{5} r_i - \thetanote{3}{5} r_{i-1} \right ]
            \left ( \avgsub{f}{i} - \avgsub{f}{i+1} \right ) }
           {\left [ \thetanote{3}{4} r_i - \thetanote{3}{4} r_{i-1} \right ]
            \left [ \thetanote{3}{5} r_i - \thetanote{3}{5} r_{i+1} \right ] -
            \left [ \thetanote{3}{4} r_i - \thetanote{3}{4} r_{i+1} \right ]
            \left [ \thetanote{3}{5} r_i - \thetanote{3}{5} r_{i-1} \right ]
          } \enskip ,
\end{equation}
\begin{equation}
a_1 = \frac{5}{3} 
      \frac{1}{\left [ \thetanote{3}{5} r_i - \thetanote{3}{5} r_{i+1} \right ]}
      \left \{ \avgsub{f}{i} - \avgsub{f}{i+1} - \frac{3}{4} a_2 
           \left [ \thetanote{3}{4} r_i - \thetanote{3}{4} r_{i+1} \right ]
     \right \} \enskip ,
\end{equation}
and
\begin{equation}
a_3 = \avgsub{f}{i} - \frac{3}{5} a_1 \thetanote{3}{5} r_i - \frac{3}{4} a_2 \thetanote{3}{4} r_i \enskip .
\end{equation}
Together with Eq. \ref{eq:reconstpoly}, these reconstruct the data in
the $i$-th zone.  When we refine, two children are created from the
$i$-th zone, each taking up half the interval.  This is illustrated in
Figure \ref{fig:sph_quad} on the lower axis.  For convenience, we use
$ic$ to refer to the child index space, with $r_{ic-1/2} = r_{i-1/2}$,
as shown in the figure.  The child data is found by integrating
Eq. \ref{eq:reconstpoly} over the child zone as
\begin{equation}
\avgsub{f}{ic} = \frac{4 \pi}{V_{ic}} \int_{r_{ic-1/2}}^{r_{ic+1/2}} r^2 dr f(r)\enskip ,
\end{equation}
where $ic$ refers to the child zone index.  Carrying out the
integration, we find the value of the child data, in terms of the
parent data and it's immediate neighbors is
\begin{equation}
\avgsub{f}{ic} = \frac{3}{5} a_1 \thetanote{3}{5} r_{ic} +
                 \frac{3}{4} a_2 \thetanote{3}{4} r_{ic} + a_3 \enskip .
\end{equation}

This procedure is guaranteed to be conservative, but not guaranteed to
be monotonic.  If monotonicity is a requirement, then we can check
whether the newly initialized child data falls outside the limits of
the coarse zones in the stencil, and if so, fall back to direct
insertion.
\section{MODEL FLAMES}
\label{sec:model-flame-appendix}

In this appendix, we describe methods and results for `model'
flames, both to connect with the combustion literature and
to provide simpler models for understanding the astrophysical
flames.

\subsection{Equation Of State}

For the simple model flames we use a ideal gas gamma-law equation of state
\begin{equation}
P = (\gamma -1 ) \rho \epsilon \enskip ,
\end{equation}
supplemented with an expression relating the temperature and the
pressure (ideal gas approximation),
\begin{equation}
P = \frac{N_a k}{\bar{A}} \rho T \enskip ,
\end{equation}
where $\epsilon$ is the internal energy / gram, $N_a$ is Avogadro's number,
$k$ is Boltzmann's constant, and $\bar{A}$ is the average atomic mass 
of the mixture of fluids.  For the model flames, we use $\gamma = 5/3$.

\subsection{Diffusion}

For some of our model flames, we allow finite $\Lewis$, and so species
diffusion is also calculated.  Here, the diffusion term of the species
evolution equation is
\begin{equation}
\label{eq:speciesdiff}
    \frac{\partial \rho X_i}{\partial t} 
    = \nabla \cdot (D_s \nabla \rho X_i),
\end{equation}
where $X_i$ is the mass fraction of species $i$, and $D_s$ is the species
diffusivity.   When species diffusion is included, $D_s$ is assumed
constant and fixed.   The species diffusion is solved in the same way
as the thermal diffusion, described in \S~\ref{sec:methods:diff}.

\subsection{Burning}

In \S~\ref{sec:methods:burning:cf88}, we described the burning
rates for the astrophysical flames.  Here we describe simpler
burning rates used for the model flames.

\subsubsection{KPP}
\label{sec:methods:burning:kpp}

A simple burning mechanism, often used by applied mathematicians because
it is tractable analytically, is due to \cite{kpp}, known for
the authors of that paper as KPP.  It is a one-step reaction
that irreversibly converts fuel to ash:
\begin{equation}
\label{eq:origkpp}
\frac{d \Theta}{dt} = -\frac{dX_f}{dt} = a \Theta X_f
\end{equation}
where $\Theta$ is a linearly scaled temperature such that $\Theta =
1$ corresponds to the temperature of the burned ash, and $\Theta = 0$
corresponds to the temperature of the unburned fuel.  The concentration
(mass fraction) of fuel is given by $X_f$, also varying from 0--1.   An
arbitrary parameter $a$ sets the rate of reaction.

Along with $a$ and the diffusivities $D_{th}$ and $D_s$, we
have another parameter we can choose --- the energy input of the
burning with respect to the background energy.  We parameterize this by
the ratio of the energy released by burning a gram of material to the
thermal energy of a gram of fuel in the unburned state
\begin{equation}
\label{eq:qdefn}
q \equiv \frac{\Delta e / (\mu_f) }{C_P T_u},
\end{equation}
where $\Delta e$ is the binding energy released by burning a single
particle of fuel, $\mu_f$ its mass, and $C_P$ is the heat capacity at
constant pressure of the fuel.

\subsubsection{Arrhenius}
\label{sec:methods:burning:arr}

Real flames, both astrophysical and terrestrial, have such strongly peaked
temperature dependencies that their structure can be
approximately be broken up into a diffusive ``preheat'' zone, where
reactions are negligible, and a thin reaction zone, where most of the
burning occurs and the temperature is roughly constant.
Model flames using a KPP burning rate lack this important feature, because
of KPP's fairly gentle temperature dependence.

Thus, as a more realistic model, we consider an Arrhenius rate (see
for example \citealt{williams})
\begin{equation}
\label{eq:arr}
\frac{d T}{dt} = a T_a \rho^n X_f^m e^{-T_a/T},
\end{equation}
where $T_a$ is an `activation temperature' large compared to any of
the temperatures in the system.    If the molecular weights of
the fuel and ash are the same, then in terms of $X_f$ this becomes
\begin{equation}
\frac{d X_f}{dt} = -a \frac{T_a}{q T_u} \rho^{n-1} X_f^m e^{-T_a/T}.
\end{equation}

\subsection{Results}

For the scale-free model flames, dimensional values of parameters
are somewhat arbitrary.   Parameters were chosen so that the fastest
outward-propagating spherical flames would have $d r_f/dt \approx 1$
in a domain where the slowest sound speed would be $c_s = 100$, and all
flames were run with these parameters.  $D_{th}$ was fixed at
$0.0625~\mathrm{cm^2 \ s^{-1}}$, and $D_s = D_{th}/\Lewis$.   The
unburned state was set to $T_u = 0.1, \rho_u=1.$ and the energy release
was set so that $T_b^0 = 0.64$.  The atomic weight of both the fuel and
ash were set to 1385.1.

We consider two sets of flames --- `KPP flames', which have an ideal
gas EOS and burn with the KPP burning rate described in
\S\ref{sec:methods:burning:kpp}, and `Arrhenius flames', which also use
the ideal gas EOS but burn with an Arrhenius law, described in
\S\ref{sec:methods:burning:arr}.  Simulations were run with varying
temperature dependence and Lewis numbers.  Simulations with finite large
Lewis numbers become computationally increasingly expensive, as an
increasingly smaller species diffusion zone must be resolved
adequately.

\subsubsection{KPP flames}

Results for KPP flames are shown in
Fig~\ref{fig:kppmarksteinkarlovitz}.   Shown are results from
simulations with $a = 0.3$, $\Lewis = 1$, and $a = 0.3$,
$\Lewis=\infty$.  Data from outwardly-propagating and
inwardly-propagating spherical flames, and planar flames are shown.
Quantitative results are summarized in
Table~\ref{table:kppflamedata}.   The resolution used was that chosen
in the Appendix for the $\Zeldovich \sim 5$, $\Lewis = \infty$ flame,
although the KPP flames are much thicker.

\begin{figure}
\plottwo{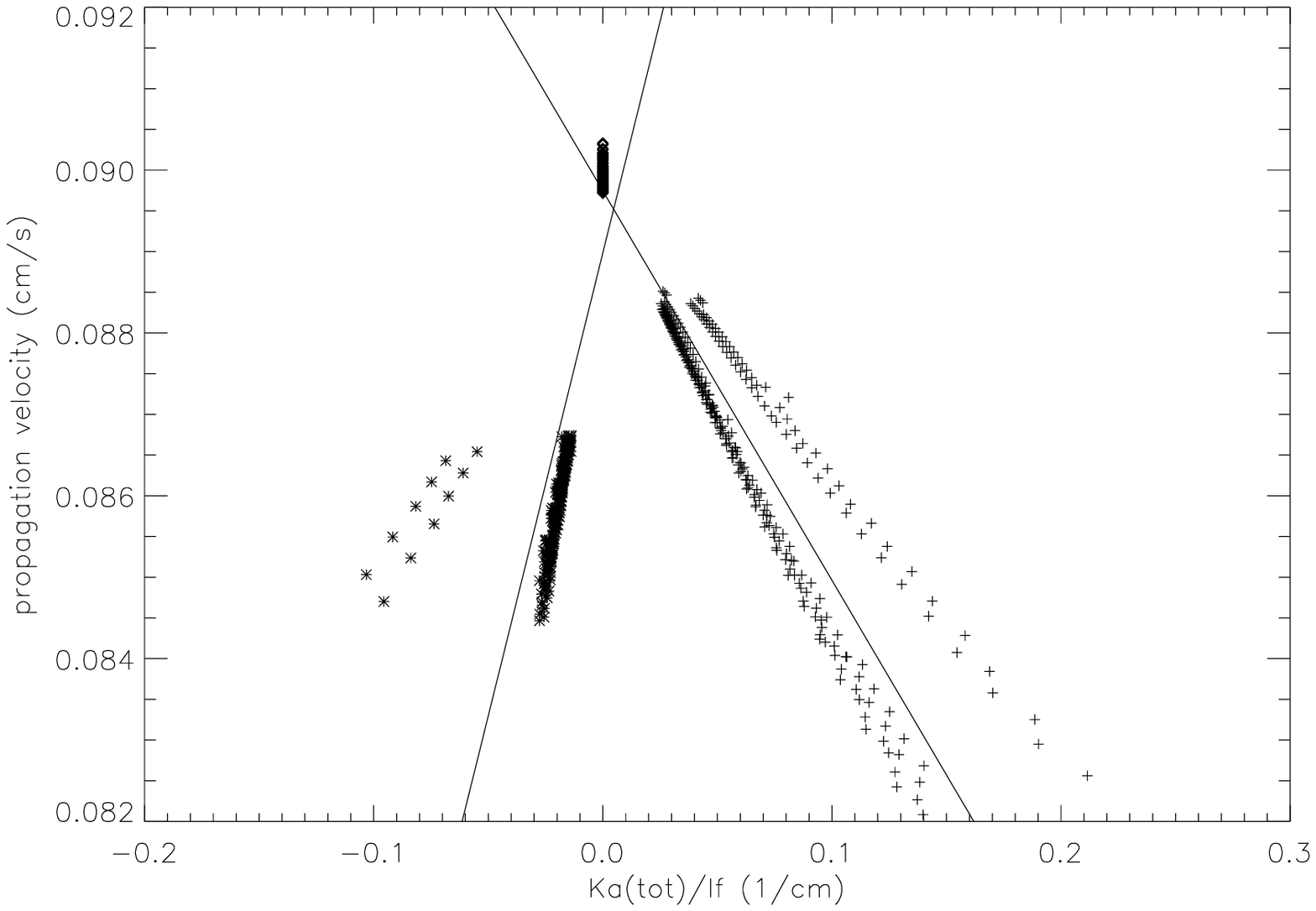}{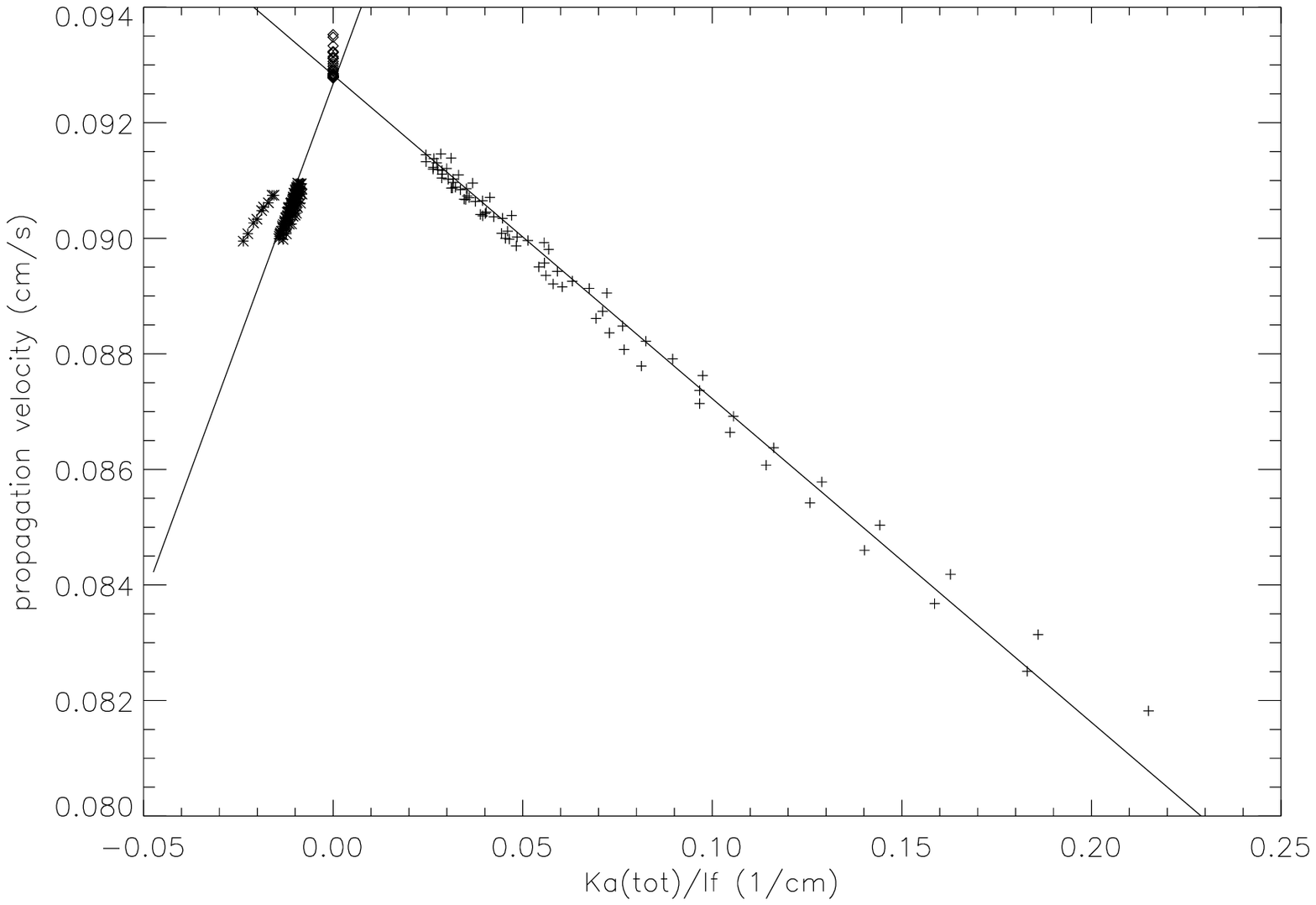}
\caption{\label{fig:kppmarksteinkarlovitz} 
         Flame speed versus 
         $\Karlovitz_T / l_f$ for KPP flames, with
         $\Lewis = 1$ (left) and $\Lewis = \infty$ (right). 
         For each graph,
         data points for positive $\Karlovitz_T/l_f$ are from simulations
         of a spherically expanding flame, data points for 
         $\Karlovitz_T/l_f = 0$ are for a planar flame, and 
         those for negative $\Karlovitz_T/l_f$ are for a spherical
         flame propagating inwards.
    }
\end{figure}

\begin{table}
\begin{center}
\begin{tabular}{cdddd@{$\, \pm \,$}ld@{$\, \pm \,$}l}
\tableline
\tableline
$\Lewis$ & \multicolumn{1}{c}{$S_l^0$ (cm s$^{-1}$)} & \multicolumn{1}{c}{$l_f^{0(I)}$ (cm)} & \multicolumn{1}{c}{$l_f^{0(II)}$ (cm)} & \multicolumn{2}{c}{$\MarksteinLength^{(+)}$ (cm)} & \multicolumn{2}{c}{$\MarksteinLength^{(-)}$ (cm)}\\
\tableline
1      & 0.0896 & 5.43  & 4.42  & -.533 & .139 & +1.28 & .60 \\
$\infty$& 0.0927 & 5.31  & 4.23  & -.604 & .139 & +1.92 & .82 \\
\tableline
\end{tabular}

\end{center}
\caption{KPP flame data.}
\label{table:kppflamedata}
\end{table}

We use the $\enuc$-based method for flame velocities as described in
\S\ref{subsec:flamevel}, even though the approximation of a thin
reaction front is a poor one for these KPP flames.  The quantitative
results shown here can be changed significantly by choosing the flame
position differently within the burning zone of the flame; the errors
quoted for $\MarksteinLength$ in Table~\ref{table:kppflamedata} come
from using the approximate $\pm 1\ \mathrm{cm}$ uncertainty in flame
position shown in Fig.~\ref{fig:position-measures}.

We see here that KPP flames do not behave in a way described by the
Markstein relation.   Although the negatively and positively strained
flames separately respond significantly and linearly to strain, the
sign of the response is different in the two cases.  Straining the KPP
flames here result in a slowing down of the flame, regardless of the
sign of the strain.

That the KPP flame responds differently to strain rates of order a
flame crossing time than other flame models is easily understood.
Where the astrophysical flames described in \S\ref{sec:theory} have a
distinct burning and preheat region, the KPP flames do not.   Thus, in
this case, the straining stretches not just the preheat zone ---
essentially the preconditioning for the burning zone --- but the
burning zone itself.   

\subsubsection{Arrhenius flames}

Arrhenius flames were run with $\Zeldovich \sim 5$ and 8.
The flame structures for $\Zeldovich \sim 5$
planar flames with different Lewis numbers are shown in
Fig.~\ref{fig:all-le-plane}; $\Lewis = \infty$ flames with positive,
zero, and negative strain are shown in
Fig.~\ref{fig:le-strain-struct}.  The plots of flame speed versus
scaled dimensionless strain are shown in
Fig.~\ref{fig:ze5marksteinkarlovitz} and
Fig.~\ref{fig:ze8marksteinkarlovitz}.  Table
\ref{table:arrheniusflamedata} gives the quantitative results from
these flames.   Quoted errors on $\MarksteinLength$ come from
uncertainties in the best fit.    The $\Lewis = \infty$, $\Zeldovich
\sim 8$ flame proved too difficult to reliably ignite with the constant
diffusivity used here.

\begin{figure}
\plotone{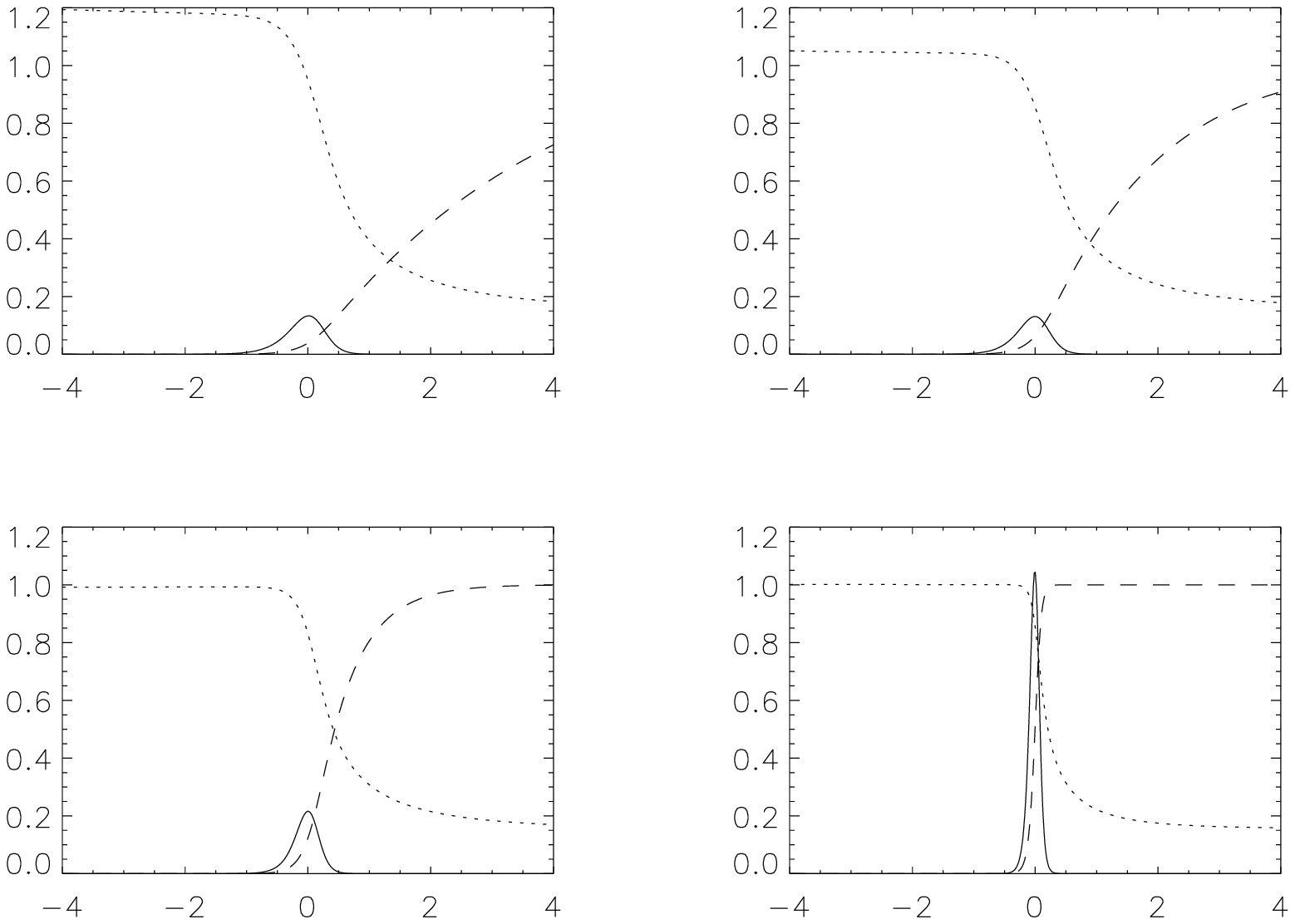}
\caption{\label{fig:all-le-plane} 
         Shown is the flame structure (solid line is scaled $\enuc$,
         dashed line is $X_f$, and dotted temperature) for planar
         Arrhenius $\Zeldovich \sim 5$ flames, with $\Lewis =
         1,2,5,\infty$ at the top left, top right, bottom left, and
         bottom right.  $\enuc$ is scaled to 81434.6 ergs g$^{-1}$
         s$^{-1}$, and temperature to the adiabatic temperature of the
         $\Lewis = \infty$ flame, $T_b^0 = 0.64$.  As the species
         diffusion decreases, the size of the location where there is
         both enough heat and fuel to burn decreases, and the burning
         zone narrows considerably, increasing peak $\enuc$ (although
         the total burning rate only changes by a factor of 2.)  X-axis is
         distance into the unburned material from the flame position
         $r_f$.  }
\end{figure}

\begin{figure}
\plotone{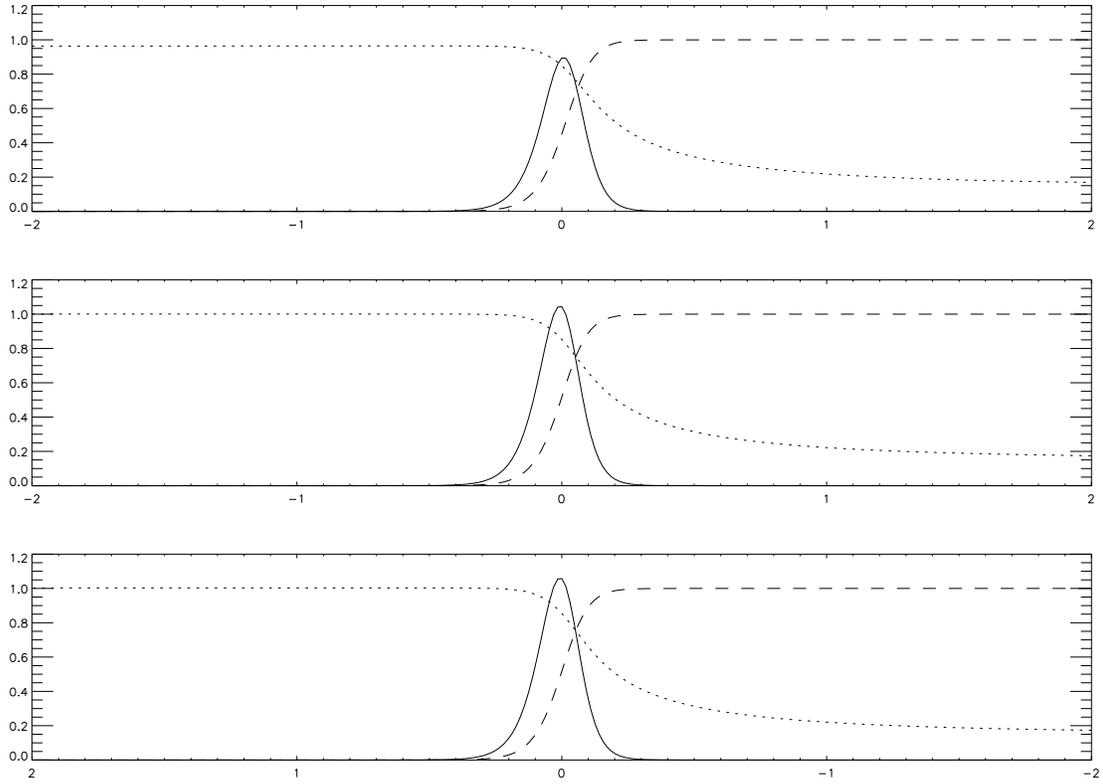}
\caption{\label{fig:le-strain-struct} 
         Shown is the flame structure (solid line is scaled $\enuc$,
         dashed line is $X_f$, and dotted temperature) for Arrhenius
         $\Zeldovich \sim 5$ flames which are the outward propagating
         (top), planar (middle), and inward propagating (bottom).
         $\Karlovitz_T / l_f$ is $+.2625 \ \mathrm{cm^{-1}}$ for the
         outward propagating flame, $0 \ \mathrm{cm^{-1}}$ for the
         planar flame, and $-.03675 \ \mathrm{cm^{-1}}$ for the inward
         flame.  Although the change in the thermal and fuel structure is
         very small, the flame speed changes by $\sim 10\%$.  }
\end{figure}

\begin{figure}
\plotone{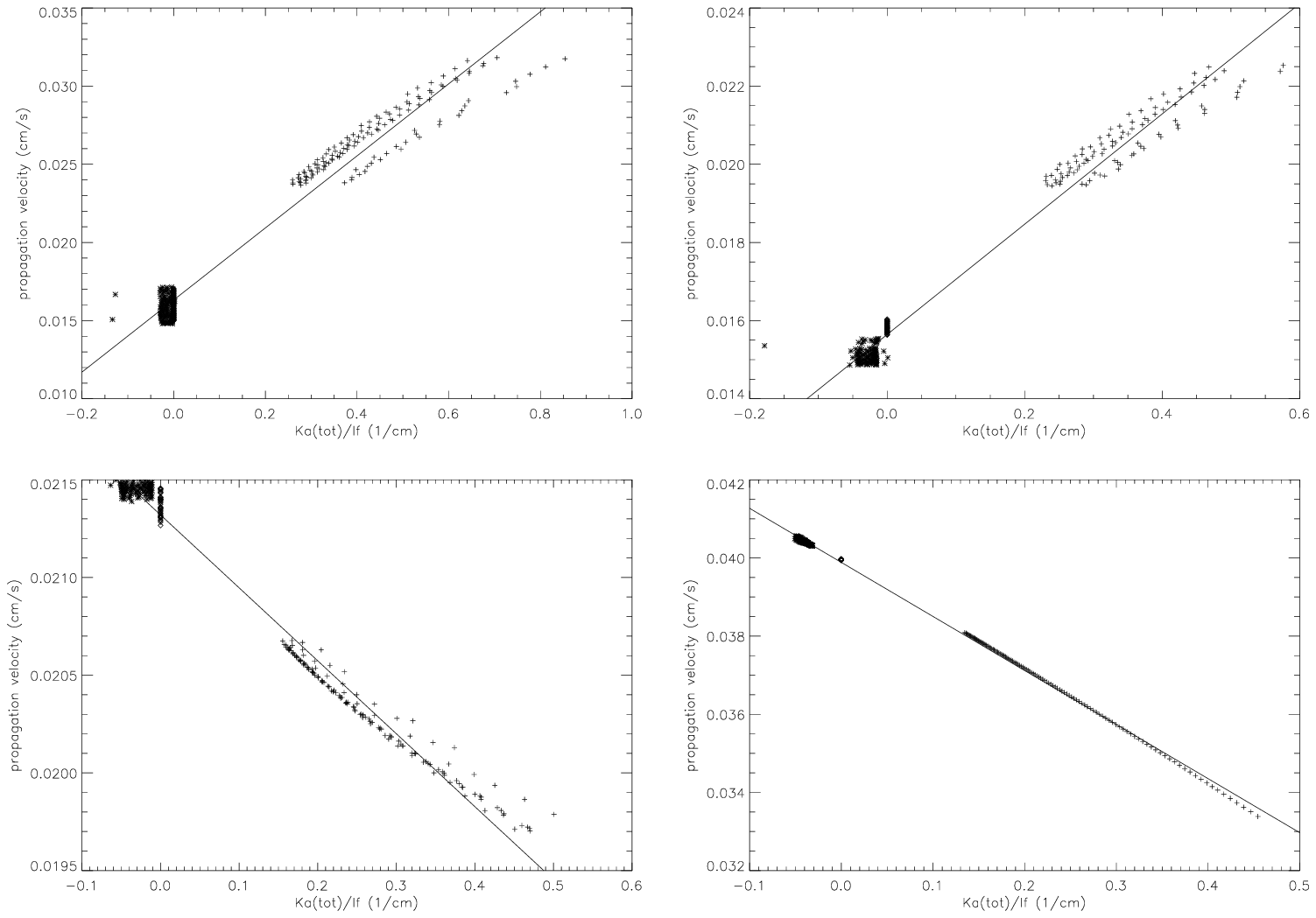}
\caption{\label{fig:ze5marksteinkarlovitz} Flame speed versus 
         $\Karlovitz_T / l_f$ for $\Zeldovich \sim 5$ Arrhenius flames.
         Shown are plots for $\Lewis = 1$ (top left),
         $\Lewis = 2$ (top right), $\Lewis = 5$ (bottom left),
         and $\Lewis = \infty$ (bottom right).  
         For each graph,
         data points for positive $\Karlovitz_T/l_f$ are from simulations
         of a spherically expanding flame, data points for 
         $\Karlovitz_T/l_f = 0$ are for a planar flame, and 
         those for negative $\Karlovitz_T/l_f$ are for a spherical
         flame propagating inwards.
    }
\end{figure}

\begin{figure}
\plottwo{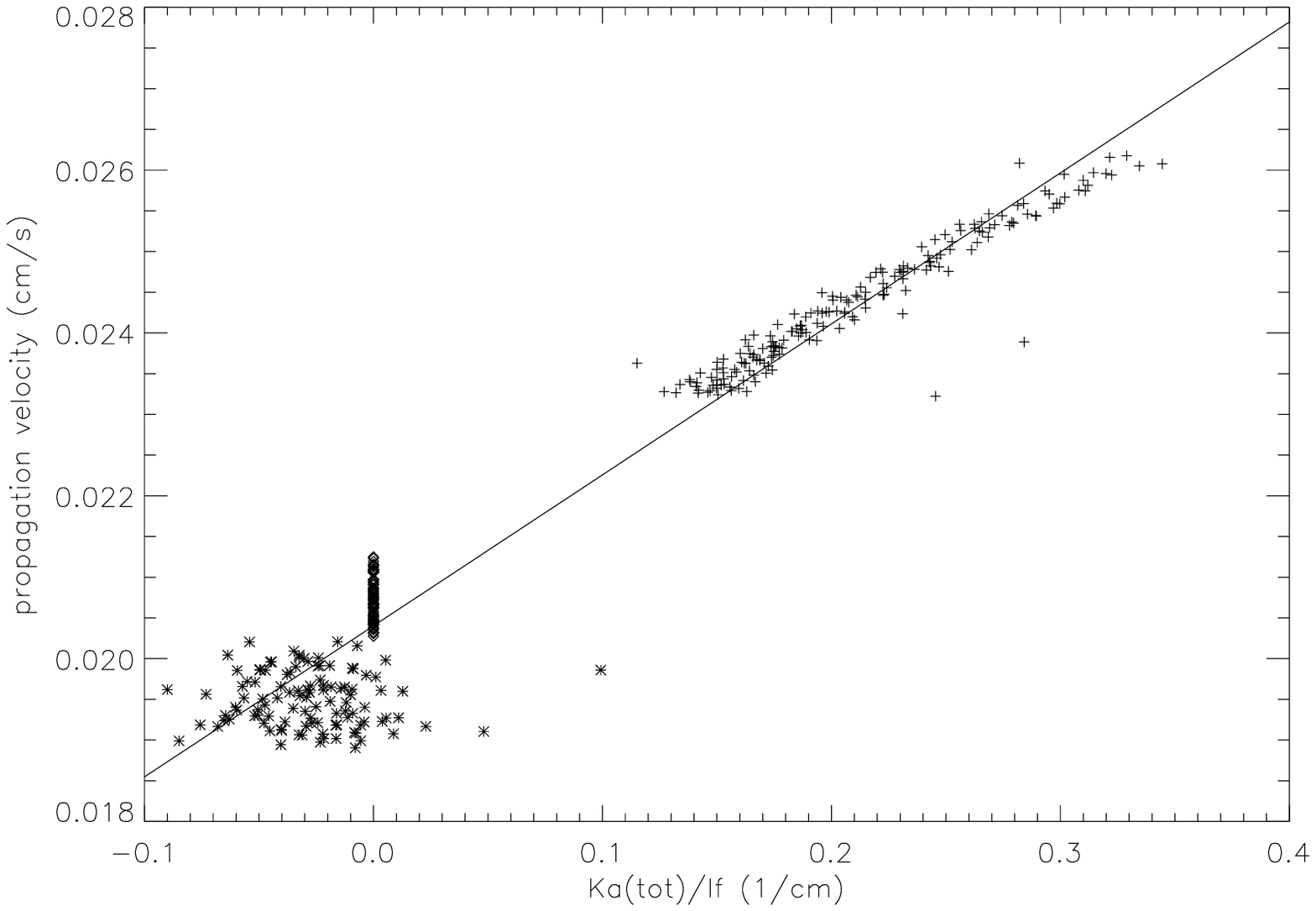}{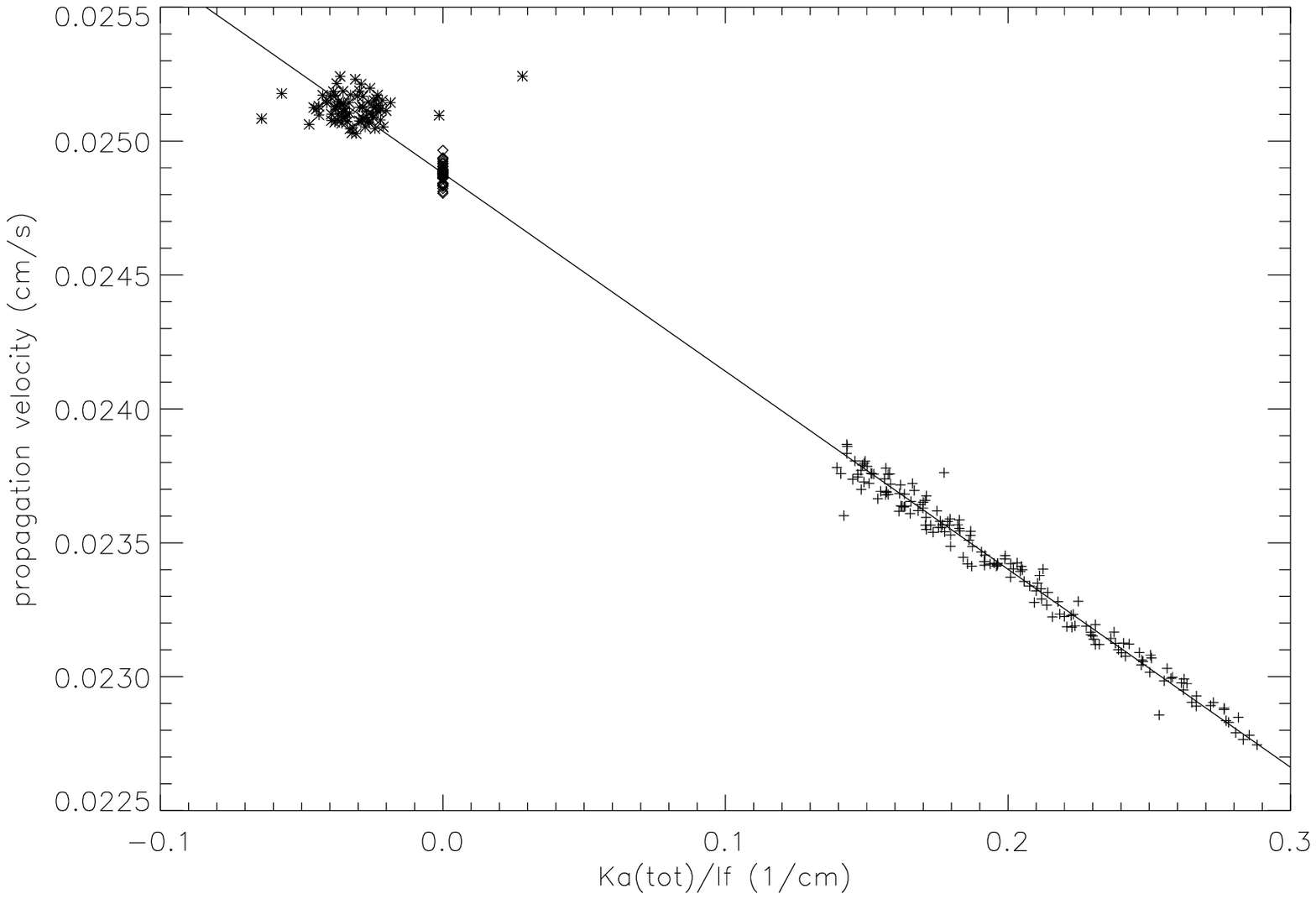}
\plotone{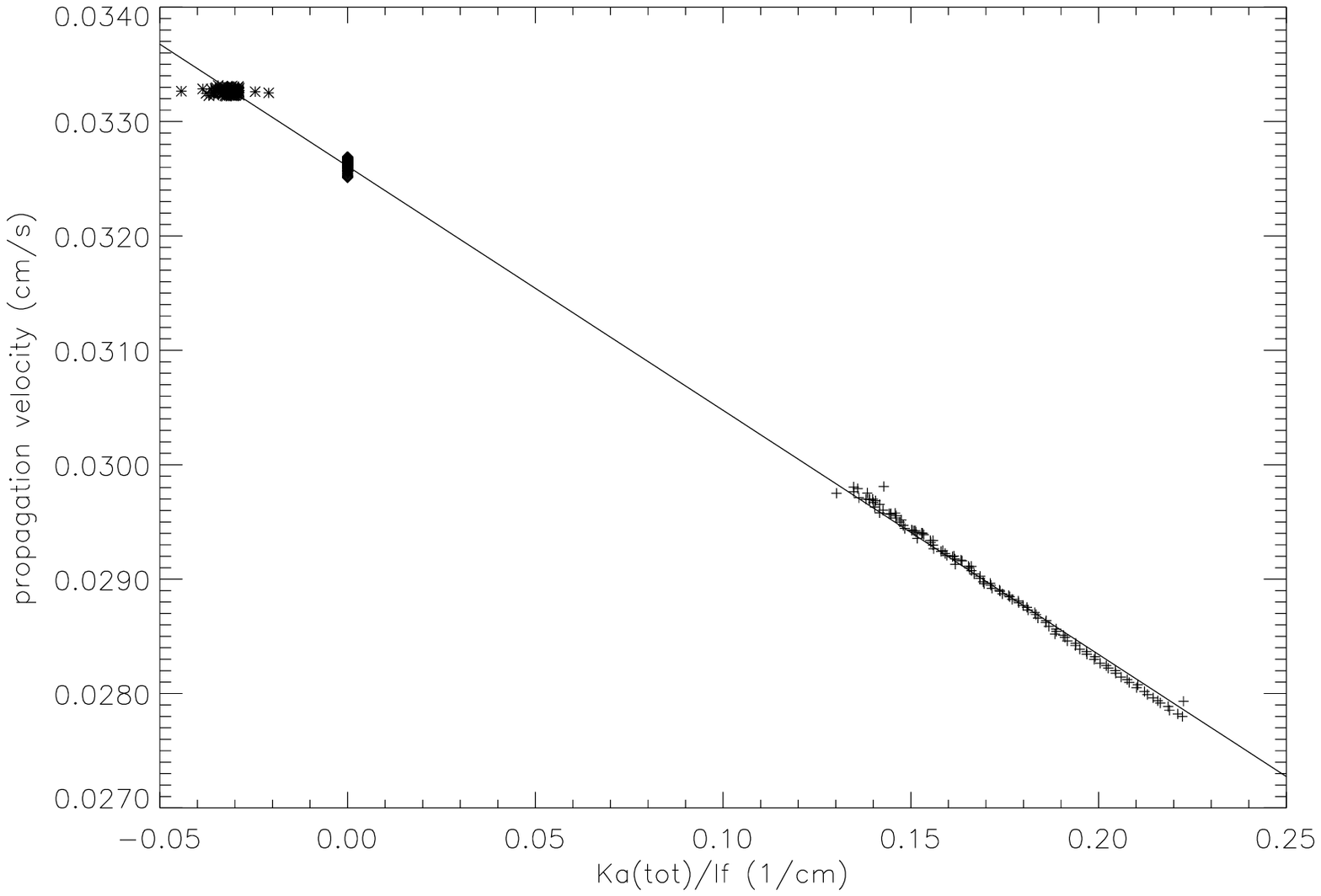}

\caption{\label{fig:ze8marksteinkarlovitz} Flame speed versus 
         $\Karlovitz_T / l_f$ for $\Zeldovich \sim 8$ Arrhenius flames.
         Shown are plots for $\Lewis = 1$ (top left), $\Lewis = 2$ (top right),
         and $\Lewis = 5$ (bottom).  For each graph,
         data points for positive $\Karlovitz_T/l_f$ are from simulations
         of a spherically expanding flame, data points for 
         $\Karlovitz_T/l_f = 0$ are for a planar flame, and 
         those for negative $\Karlovitz_T/l_f$ are for a spherical
         flame propagating inwards.
    }
\end{figure}

\begin{table}
\begin{center}

\begin{tabular}{ccrddd@{$\, \pm \,$}ldd}
\tableline
\tableline
$\Zeldovich$ & $\Lewis$ & $S_l^0$ (cm s$^{-1}$) & \multicolumn{1}{c}{$l_f^{0(I)}$ (cm)} & \multicolumn{1}{c}{$l_f^{0(II)}$ (cm)} & \multicolumn{2}{c}{$\MarksteinLength$ (cm)} & \multicolumn{1}{c}{$\Markstein ^{(I)}$} & \multicolumn{1}{c}{$\Markstein ^{(II)}$} \\
\tableline
5.37 & 1      & 0.0227 & 1.547  & 0.827  & +0.452 & 0.003 & +.292 & +.546 \\
     & 2      & 0.0269 & 1.344  & 0.694  & -0.195 & 0.001 & -.150 & -.281 \\
     & 5      & 0.0330 & 1.078  & 0.5528 & -0.434 & 0.002 & -.403 & -.823 \\
     &$\infty$& 0.0399 & 0.8906 & 0.4233 & -0.347 & 0.001 & -.334 & -.702 \\
\tableline
8.06 & 1      & 0.0204 & 1.636 & 1.481  & +0.909  & 0.002 & +0.555  &  +0.614 \\
     & 2      & 0.0248 & 1.326 & 0.5422 & -0.297  & 0.001 & -0.224  &  -0.548 \\
     & 5      & 0.0326 & 1.025 & 0.4052 & -0.654  & 0.001 & -0.638  &  -1.614 \\
\tableline
\end{tabular}

\end{center}
\caption{Arrhenius flame data.}
\label{table:arrheniusflamedata}
\end{table}

The Arrhenius flames response to strain is linear, and the Markstein
lengths are of order the flame thickness or smaller.   The sign of the
response is negative for $\Lewis = \infty$ flames, as expected.  The sign
changes for smaller Lewis number as the effect of fuel diffusion into
the ash, which acts in the opposite direction of the thermal diffusion
into the fuel, becomes more significant.

One sees in these results and in the astrophysical results that flame
response to negative curvature is noisier than that to positive curvature.
This is caused by the physical setup of the simulations.  For the ingoing
flames, pressure waves caused by ignition transients can be trapped inside
the spherical flame due to the density jump at the flame's position.
In our 1-d simulations, the pressure waves bounce along the interior of
the flame.  The waves bounce off of the reflecting boundary condition,
correctly modelling the same wave coming from the part of the spherical
domain on the other side of the origin.   The pressure waves are only
amplified as the flame moves inwards.    This is shown on the left of
Fig.~\ref{fig:preswaves-noise} for the $\Lewis = 5$, $\Zeldovich \sim 5$
ingoing Arrhenius flame.   On the right of Fig.~\ref{fig:preswaves-noise}
is plotted the same values for the corresponding outward-propagating
flame.   Here, the transient pressure waves can largely leave the domain.

\begin{figure}
\plottwo{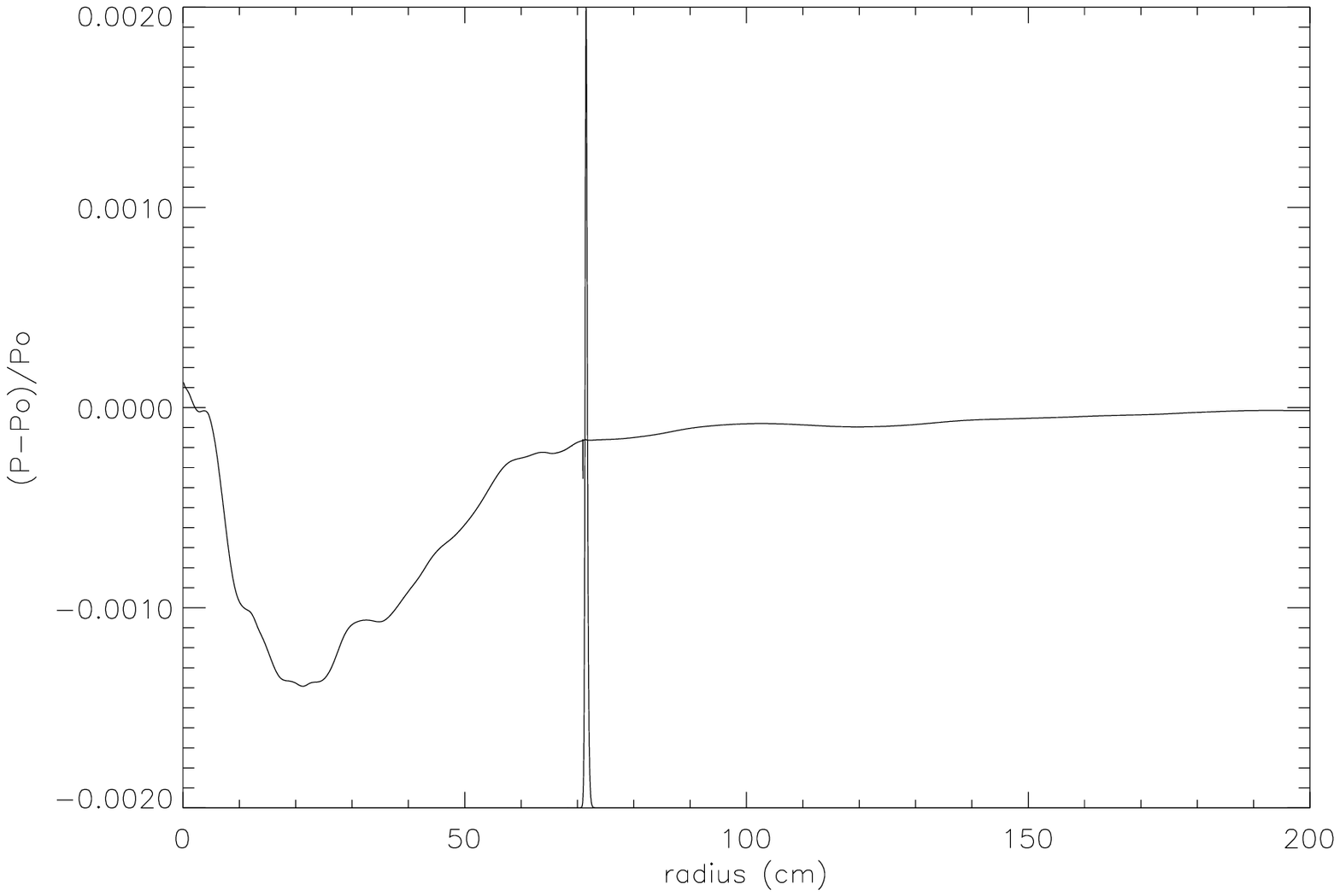}{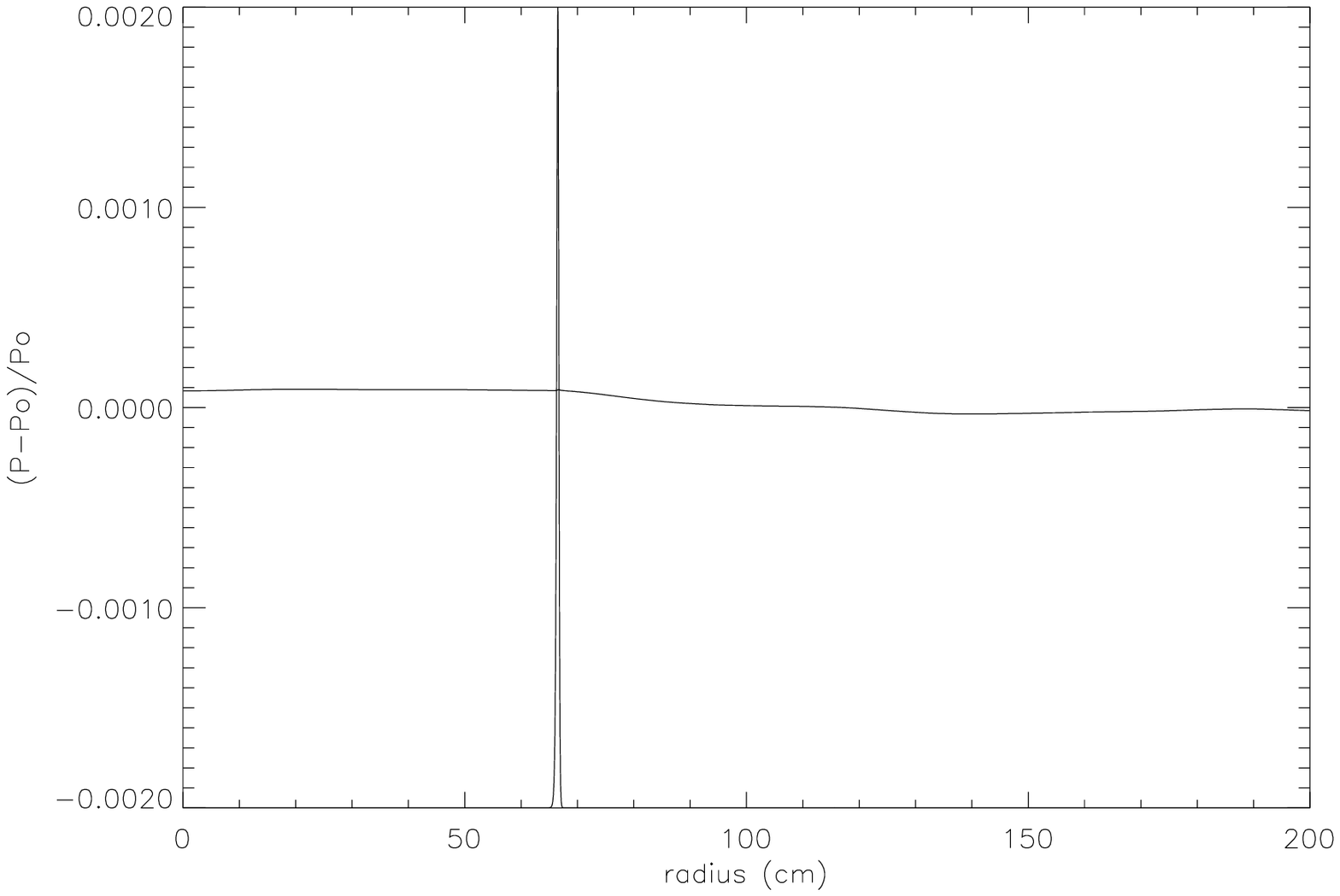}
\caption{\label{fig:preswaves-noise}
        Scaled pressure fluctuation ($(P-P_0)/P_0$) in the computational
	domain, with scaled $\enuc$ to show the flame's position, for
        an inward-propagating $\Zeldovich \sim 5$, $\Lewis = 5$ Arrhenius
        flame on the left, and the corresponding outward-propagating 
        flame on the right.   Pressure waves can be trapped inside the
        flame (eg., in the unburned fuel) for the inward flame, and amplified
        as the flame moves inwards; these fluctuations in the fuel state
        lead to the observed scatter in burning rates.    The pressure
        waves are largely able to leave the domain in the outward-propagating
        case, causing reduced scatter.
    }
\end{figure}

The pressure fluctuations, while small ($\approx 0.2\%$ for the ingoing
flames, and $\approx 0.02\%$ for the outgoing flames), are enough to
slightly modify the local burning, leading to the observed scatter in
measured burning velocities.   As the temperature sensitivity of the
flame increases, the noise caused by the same pressure fluctuations become
larger; this explains the increased noise in the $\Zeldovich \sim 8$
flames.







\clearpage

%
%
%

\clearpage

\bibliographystyle{apj}
\bibliography{flamecurve}

\end{document}